 \documentclass[tradiabstract]{aa} 

\usepackage{graphicx,txfonts}
\usepackage{amsmath}
\usepackage{mathtools}
\usepackage{relsize}

\newcommand{\Lsun} {L$_{\odot}$}
\newcommand{\Msun} {M$_{\odot}$}

\begin{document}

\title{Dust in the diffuse interstellar medium}

\subtitle{Extinction, emission, linear and circular
  polarisation\thanks {Based on ESO: 386.C-0104}}

\titlerunning{Dust in the diffuse ISM}

\author {R.~Siebenmorgen\inst{1}, N.V. Voshchinnikov\inst{2,3}, S. Bagnulo\inst{4}}
\institute{European Southern Observatory, Karl-Schwarzschild-Str. 2, D-85748
  Garching b. M\"unchen, Germany; email: Ralf.Siebenmorgen@eso.org
\and
Sobolev Astronomical Institute,
St.~Petersburg University, Universitetskii prosp. 28,
           St.~Petersburg, 198504 Russia,
\and
 Isaac Newton Institute of Chile, St.~Petersburg Branch
\and
Armagh Observatory, College Hill, Armagh BT61 9DG, UK
}

\date{Received April 16, 2013 / Accepted xxx, xxx}

\abstract{We present a model for the diffuse interstellar dust that
  explains the observed wavelength-dependence of extinction, emission,
  linear and circular polarisation of light. The model is set-up
      with a small number of parameters. It consists of a mixture of
  amorphous carbon and silicate grains with sizes from the molecular
  domain of 0.5 up to about 500\,nm.  Dust grains with radii larger
  than 6\,nm are spheroids.  Spheroidal dust particles have a
      factor 1.5 -- 3 larger absorption cross section in the far IR
      than spherical grains of the same volume. Mass estimates derived
      from submillimeter observations that ignore this effect are
      overestimated by the same amount. In the presence of a
  magnetic field, spheroids may be partly aligned and polarise
  light. We find that polarisation spectra help to determine the upper
  particle radius of the otherwise rather unconstrained dust size
  distribution.  Stochastically heated small grains of graphite,
  silicates and polycyclic aromatic hydrocarbons (PAHs) are
  included. We tabulate parameters for PAH emission bands in various
  environments. They show a trend with the hardness of the radiation
  field that can be explained by the ionisation state or hydrogenation
  coverage of the molecules.  For each dust component its relative
  weight is specified, so that absolute element abundances are not
  direct input parameters.  The model is confronted with the average
  properties of the Milky Way, which seems to represent dust in the
  solar neighbourhood.  It is then applied to specific sight lines
  towards four particular stars one of them is located in the
      reflection nebula NGC\,2023. For these sight lines, we present
  ultra-high signal-to-noise linear and circular spectro-polarimetric
  observations obtained with FORS at the VLT.  Using prolate rather
  than oblate grains gives a better fit to observed spectra; the axial
  ratio of the spheroids is typically two and aligned silicates are
  the dominant contributor to the polarisation.}  

\maketitle

\section{Introduction}

Interstellar dust grains absorb, scatter, and polarise radiation and
emit the absorbed radiation through non-thermal and thermal
mechanisms.  Dust grains not only absorb and scatter stellar photons,
but also the radiation from dust and gas. In addition, interstellar
dust in the diffuse interstellar medium (ISM) and in other
environments that are illuminated by UV photons show photoluminescence
in the red part of the spectrum, a contribution known as Extended Red
Emission (ERE, Cohen et al. 1975; Witt et al. 1984).

Clues to the composition of interstellar dust come from observed
elemental depletions in the gas phase. It is generally assumed, that
the abundances of the chemical elements in the ISM are similar to
those measured in the photosphere of the Sun. The abundances of the
elements of the interstellar dust (the condensed phase of matter) are
estimated as the difference between the elemental abundances in the
solar photosphere (Asplund et al. 2009) and that of the
gas-phase. Absolute values of the interstellar gas-phase abundance of
element [X] are given with respect to that of hydrogen [H]. Such
[X]/[H] ratios are often derived from the analysis of absorption
lines.  (Note that oscillator strengths of some species e.g., CII
2325\AA \/ are uncertain up to a factor of 2, see discussion in Draine
2011).  The most abundant condensible elements in the ISM are C, O,
Mg, Si and Fe. When compared to the values of the Sun, elements such
as Mg, Si and Fe, which form silicates, in the gas-phase appear
under--abundant. By contrast, oxygen represents a striking exception,
as it appears over--abundant towards certain sight-lines
(Voshchinnikov \& Henning 2010).  Another important dust
forming element, C, cannot be characterised in detail because it has
been analysed only in a restricted number of sight-lines, leading in
some cases to inaccurate values of its abundance (Jenkins 2009);
Parvahti et al. 2012). It is widely accepted that cosmic dust
consists of some form of silicate and carbon.

Stronger constraints on the composition of interstellar grains come
from the analysis of their spectroscopic absorption and emission
signatures. The observed extinction curves display various spectral
bands.  The most prominent ones are the 2175\,\AA\ bump, where
graphite and polycyclic aromatic hydrocarbons (PAHs) have strong
electronic transitions, and the 9.7\,$\mu$m and 18\,$\mu$m features
assigned to Si-O stretching and O-Si-O bending modes of silicate
minerals, respectively.  In addition, there are numerous weaker
features, such as the 3.4\,$\mu$m absorption assigned to C-H
stretching modes (Mennella et al. 2003), and the diffuse interstellar
bands in the optical (Krelowski 2002). The observed 9.7 and 18\,$\mu$m
band profiles can be better reproduced in the laboratory by amorphous
silicates than crystalline structures. Features that are assigned to
crystalline silicates, such as olivine (Mg$_{2x}$ Fe$_{2-2x}$
SiO$_{4}$ with $x \sim 0.8$), have been detected in AGB and T Tauri
stars and in comet Hale-Bopp (see Henning 2010 for a review). However,
since these features are not seen in the ISM, the fraction of
crystalline silicates in the ISM is estimated $\la 2$\,\% (Min et
al. 2007).  Dust in the diffuse ISM appears free of ices that
are detected in regions shielded by $A_{\rm V} > 1.6$\,mag (Bouwman et
al. 2011, Pontoppidan et al. 2003, Siebenmorgen \& Gredel 1997,
Whittet et al. 1988).

In the IR, there are conspicuous emission bands at 3.3, 6.2, 7.7, 8.6,
11.3 and 12.7\,$\mu$m, as well as a wealth of weaker bands in the 12
-- 18\,$\mu$m region. These bands are ascribed to vibrational
transitions in PAH molecules. PAHs are planar structures, and consist
of benzol rings with hydrogen attached. Less perfect structures may be
present where H atoms are replaced by OH or CN and some of the C atoms
by N atoms (Hudgins et al. 2005). PAHs may be ionised: PAH$^+$ cations
may be created by stellar photons, and PAH$^-$ anions may be created
by collisions of neutral PAHs with free e$^-$.  The ionisation degree
of PAH has little influence on the central wavelength of the emission
bands, but has a large impact on the feature strengths (Allamandola et
al. 1999, Galliano et al. 2008). Feature strengths also depend on the
hardness of the exciting radiation field, or on the hydrogenation
coverage of the molecules (Siebenmorgen \& Heymann 2012).

The extinction curve gives the dust extinction as a function of
wavelength.  It provides important constraints on dust models, and in
particular on the size distribution of the grains.  Important work has
been published by Mathis et al. (1977), who introduced their
so--called MRN size distribution, and by Greenberg (1978), who have
presented his grain core--mantle model.  Another important constraint
on dust models is provided by the IR emission. IRAS data revealed
stronger than expected 12 and 25\,$\mu$m emission from interstellar
clouds (Boulanger et al. 1985).  At the same time, various PAH
emission bands have been detected (Allamandola et al. 1985, 1989,
Puget \& L\'{e}ger 1989). Both emission components can only be
explained by considering dust particles that are small enough to be
stochastically heated.  A step forward was taken in the dust models by
D\'{e}sert et al. (1990), Siebenmorgen \& Kr\"ugel (1992), Dwek et
al. (1997), Li \& Draine (2001), and Draine \& Li (2007). In these
models, very small grains and PAHs are treated as an essential grain
component beside the (so far) standard carbon and silicate mixture of
large grains.

In all these studies, large grains are assumed to be spherical and,
the particle shape is generally not further discussed.  However, in
order to account for the widely observed interstellar polarisation,
non--spherical dust particles partially aligned by some mechanism need
to be considered.  This has been done e.g. by Hong \& Greenberg
(1980), Voshchinnikov (1989), Li \& Greenberg (1997) who considered
infinite cylinders. More realistic particle shapes such as spheroids
have been considered recently. The derivation of cross sections of
spheroids is an elaborate task, and computer codes have been made
available by Mishchenko (2000) and Voshchinnikov \& Farafonov (1993).
The influence of the type of spheroidal grain on the polarisation is
discussed by Voshchinnikov (2004). Dust models considering spheroidal
particles that fit the average Galactic extinction and polarization
curves were presented by Gupta et al. (2005), Draine \& Allaf-Akbari
(2006), Draine \& Fraisse (2009). The observed interstellar extinction
and polarization curves towards particular sight lines are modelled by
Voshchinnikov \& Das (2008) and Das et al. (2010).

In this paper we present a dust model for the diffuse ISM that
accounts for observations of elemental abundances, extinction,
emission, and interstellar polarisation by grains.  The ERE is not
polarised and is not further studied in this work (see Witt \& Vijh
2004 for a review). We also do not discuss diffuse interstellar bands
as their origin remains unclear (Snow \& Destree, 2011) nor various
dust absorption features observed in denser regions.  We first
describe the light scattering and alignment of homogeneous spheroidal
dust particles and discuss the absorption properties of PAHs. Then we
present our dust model. It is first applied to the observed average
extinction and polarisation data of the ISM, and the dust emission at
high Galactic latitudes. We present dust models towards four stars for
which extinction and IR data are available, and for which we have
obtained new spectro--polarimetric observations with the FORS
instrument of the VLT. We conclude with a summary of the main findings
of this work.

\section{Model of the interstellar dust}
\subsection{Basic definitions}
Light scattering by dust is a process in which an incident
electromagnetic field is scattered into a new direction after
interaction with a dust particle. The directions of the propagation of
the incident wave and the scattered wave define the so--called {\it
  scattering plane}. In this paper we define the Stokes parameters
{\it I, Q, U, V} as in Shurcliff (1962), adopting as a reference
direction the one perpendicular to the scattering plane. This way, the
reduced Stokes parameter $P_Q = Q/I$ is given by the difference
between the flux perpendicular to the scattering plane, $F^{\bot}$,
minus the flux parallel to that plane, $F^{\Vert}$, divided by the sum
of the two fluxes. In the context of this paper, Stokes $U$ is always
identically zero, so that $P_Q$ is also the total fraction of linear
polarisation $p$.

\noindent
We now define $\tau^{\Vert}$ and $\tau^{\bot}$ as the extinction
coefficients in the two directions, and $\tau_{\rm eff} =
(\tau^{\Vert} + \tau^{\bot}) /2$. In case of weak extinction
($\tau_{\rm eff} \ll 1$),  and since 
$\vert \tau^{\Vert} - \tau^{\bot} \vert \ll 1$,  the polarisation by
dichroic absorption is approximated by 
\begin{equation}
  p = {{F^{\bot} - F^{\Vert}} \over  {F^{\bot} + F^{\Vert}} }  
  = {{e^{\tau^{\Vert}} - e^{\tau^{\bot}}} \over  {e^{\tau^{\bot}} + e^{\tau^{\Vert}}} } \simeq
 \frac{\tau^{\Vert} - \tau^{\bot}}{2} \,.
\label{linpol.eq}
\end{equation}

A medium is birefringent if its refractive index depends on the
direction of the wave propagation. In this case, the phase velocity of
the radiation also depends on the wave direction, and the medium
introduces a phase retardation between two perpendicular components of
the radiation, transforming linear into circular polarisation.  In the
ISM, a first scattering event in cloud 1 will linearly polarise the
incoming (unpolarised) radiation, and a second scattering event in
cloud 2 will transform part of this linear polarisation into circular
polarisation.  Denoting by $p(1)$ the fraction of linear polarised
induced during the first scattering event, and by $N_d(2)$ the dust
column density in cloud 2, and with $\Psi$ the difference of
positional angles of polarisation in clouds 1 and 2, the circular
polarisation $ p_{\rm c}$ is a second order effect and given by

\begin{equation}
 p_{\rm c} = \frac{V}{I} = N_d(2) C_{c}(2) \ \times \ p(1) \sin(2\Psi) \,,
\label{cp.eq}
\end{equation}
where $C_{c}$ is the cross section of circular polarisation
(birefringence) calculated as the difference in phase lags introduced
by cloud 2.  Light propagation in a medium in which the grain
alignment changes is discussed by Martin (1974) and Clarke
(2010).  The wavelength dependence and degree of circular
polarisation is discussed by Kr\"ugel (2003) and that of single light
scattering by asymmetric particles by Guirado et al. (2007).

\subsection{Spheroidal grain shape \label{spheroids.sec}}

The phenomenon of interstellar polarisation cannot be explained by
spherical dust particles consisting of optically isotropic materials.
Therefore, as a simple representation of finite sized grains, we
consider spheroids. The shape of spheroids is characterised by the
ratio $a/b$ between major and minor semi-axes. There are two types of
spheroids: \textit{prolates}, such as needles, which are
mathematically described by rotation about the major axis of an
ellipse, and \textit{oblates}, such as pancakes, obtained from the
rotation of an ellipse about its minor axis. In our notation, the
volume of a prolate is the same of a sphere with radius $r = (a \cdot
b^2)^{1/3}$, and the volume of an oblate is that of a sphere with
radius $r = (a^2 \cdot b)^{1/3}$.

The extinction optical thickness, which is due to absorption plus
scattering by grains of radius $r$, is given by

\begin {equation}
\tau ({\nu}) = N_{\rm d} \  C_{\rm ext}({\nu})
\label{tauCext.eq}
\end {equation}

\noindent
and similar the linear polarisation by

\begin {equation}
p(\nu) = N_{\rm d}  \ C_{\rm p}({\nu})\,,
\label{pCp.eq}
\end {equation}

\noindent
where $N_{\rm d}$ is the total column density of the dust grains along
the line of sight and $C_{\rm {ext, p}}$ are the extinction and
linear polarisation cross sections.

\subsection{Dust alignment}

Stellar radiation may be polarised by partially aligned spheroidal
dust grains that wobble and rotate about the axis of greatest moment
of inertia. The question on how grain alignment works is not
settled. Various mechanisms such as magnetic or radiation alignment
are suggested, see Voshchinnikov (2012) for a review. We consider
grain alignment along the magnetic field $\vec{B}$ that is induced by
paramagnetic relaxation of particles having Fe impurities.  This
so--called imperfect Davis-Greenstein (IDG) orientation of spheroids
can be described by

\begin{figure}
\centering
   \includegraphics[width=8cm]{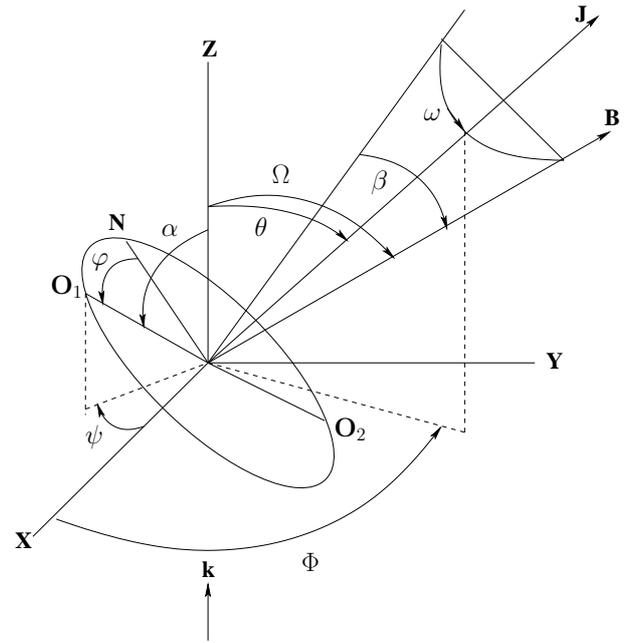}
   \caption{Geometrical configuration of a spinning and wobbling
     prolate spheroidal grain with notation by Das et al. (2010).  The
     major axis O$_1$O$_2$ of the particle spheroid is placed in the
     spinning plane NO$_1$O$_2$ that is perpendicular to the angular
     momentum $\vec{J}$.  Direction of the light propagation $\vec{k}$
     is set parallel to the $Z$-axis. We measure from $Z$ the angle $0
     \leq \Omega \leq 90\degr$ to the magnetic field $\vec{B}$, the
     angle $\alpha$ to the major rotation axis of the particle and the
     angle $\theta$ to the angular momentum $\vec{J}$; $\varphi$ is
     the spin angle, $\beta$ is the precession-cone angle, and
     $\omega$ the current precession angle.\label{angles} }
\end{figure}

\begin{equation}
{f}(\xi, \beta) = \frac{\xi \sin \beta}{(\xi^2 \cos^2 \beta  + \sin^2 \beta)^{3/2} } \,,
\end{equation}
where $\beta$ is the precession-cone angle defined in Fig.~1, and 
\begin{equation}
\xi^2  = \frac{r +
\delta_0\ (T_{\rm d}/T_{\rm g})}{r +\delta_0}\,.
\label{align}
\end{equation}

\noindent The alignment parameter, $\xi$, depends on the size of the
particle, $r$. The parameter $\delta_0$ is related to the magnetic
susceptibility of the grain, its angular velocity and temperature, the
field strength and gas temperature (Hong \& Greenberg 1980).
Voshchinnikov \& Das (2008) show that the maximum value of the
polarisation depends on $\delta_0$, whereas the spectral shape of the
polarisation does not. Das et al. (2010) are able to fit polarisation
data by varying the size distribution of the dust particles and by
assuming different alignment functions. We simplify matters and choose
$\delta_0=10\mu$m and $T_{\rm g} = 10 \ T_{\rm d}$. If the grains are
not aligned ($\xi=1$) then ${f}(\xi, \beta)=\sin \beta$; in the case
of perfect rotational alignment $\xi=0$. In the IDG mechanism
(Eq.~\ref{align}) smaller grains are better aligned than larger ones.

\subsection{Cross sections of spheroids \label{C.sec}}
Cross sections of spinning spheroids change periodically. We compute
the average cross section of spinning particles following Das et
al. (2010).  Such mean extinction ${C}_{\rm ext}$, linear $ {C}_{\rm
  p}$ and circular polarisation ${C}_{{\rm c}}$ cross sections of a
single-sized homogeneous spheroidal particle are obtained at a given
frequency $\nu$ by:
\begin{equation}
  {C}_{{\rm ext}}(\nu) =  \frac{2}{\pi} 
  \int  (Q_{{\rm ext}}^{\rm{TM}} + Q_{{\rm ext}}^{\rm{TE}}) \, r^2 \,
  f(\xi, \beta) \, \mathrm{d}{\varphi} \, \mathrm{d}{\omega} \, \mathrm{d}{\beta}\,,
\label{Cext.eq}
\end{equation}
\begin{equation}
  {C}_{{\rm p}}(\nu) = \frac{1}{\pi}  \int
 (Q_{{\rm ext}}^{\rm{TM}} - Q_{{\rm ext}}^{\rm{TE}}) \, r^2 \,
  f(\xi, \beta) \, \cos (2{\psi}) \, 
  \mathrm{d}{\varphi} \, \mathrm{d}{\omega} \,  \mathrm{d}{\beta}\,,
\label{Cp.eq}
\end{equation}
\begin{equation}
  {C}_{{\rm c}}(\nu) = \frac{1}{\pi}  \int  
(Q_{{\rm pha}}^{\rm{TM}} - Q_{{\rm pha}}^{\rm{TE}})  r^2 f(\xi, \beta) \cos (2{\psi}) \,
  \mathrm{d}{\varphi} \, \mathrm{d}{\omega} \,  \mathrm{d}{\beta}\,.
\label{Cc.eq}
\end{equation}

\noindent
Angles $\psi ,\, \varphi ,\, \omega ,\, \beta$ are shown in
Fig.~\ref{angles}.  The relations between them are defined by Hong \&
Greenberg (1980).  The efficiency factors $Q$ in Eqs.~(\ref{Cext.eq}
-- \ref{Cc.eq}), with suffix TM for the transverse magnetic and TE for
transverse electric modes of polarisation (Bohren \& Huffman 1983),
are defined as the ratios of the cross sections to the geometrical
cross-section of the equal volume spheres, $Q = C/\pi r^{2}$.  The
extinction and scattering efficiencies $Q_{{\rm ext}}$, $Q_{{\rm
    sca}}$ and phase lags $Q_{{\rm pha}}$ of the two polarisation
directions are computed with the program code provided by
Voshchinnikov \& Farafonov (1993).  The average absorption and
scattering cross sections ${C}_{{\rm abs, sca}}$ are obtained similar
to Eq.~(\ref{Cext.eq}) utilizing $Q_{{\rm abs, sca}}$, respectively.


\begin{table*}[h!tb]
\begin{center}
\caption {Band parameters of astronomical PAHs.}
\label{pah.tab}
 \begin{tabular}{|c|c|c|c|c|c|c|c|l|}
\hline
\hline
     \multicolumn{2}{|c|}{}     &      \multicolumn{3}{|c|}{} &     \multicolumn{3}{|c|}{} &  \\
     \multicolumn{2}{|c|}
{Center Wavelength} & \multicolumn{3}{|c|}{Damping Constant} & \multicolumn{3}{|c|}{Integrated Cross Section$^{a}$}  & Mode$^{b}$ \\
    \multicolumn{2}{|c|}{$\lambda_0$ ($\mu$m)}     & \multicolumn{3}{|c|}{
 $\gamma$ ($10^{12}$s$^{-1}$)} &  \multicolumn{3}{|c|}{ $ \sigma_{\rm int}$ ($10^{-22}$cm$^2$$\mu$m)}  &   \\
\hline
& & & & & & & & \\
ISM + & NGC\,2023  & Starbursts$^c$    & ISM$^d$& NGC\,2023$^d$ & Starbursts$^c$ & ISM$^d$& NGC\,2023$^d$  & \\
Starbursts& & & & & & & & \\
\hline
& & & & & & & & \\
0.2175 & & $\cdots$     &1800 &1800& $\cdots$ &8000&8000&    $\pi^* \leftarrow \pi$ transitions \\
3.3  & $\cdots$ & 20    &20 &20& 10  &20  &20& C-H stretch        \\
5.1  & $\cdots$ & 12    &20 &20& 1   &2.7   &2.7& C-C vibration        \\
6.2  & $\cdots$ & 14     &14 &14& 21  &10  &30& C-C vibration        \\
7.0  & $\cdots$ & 6    &6  &5& 13  &5  &10& C-H?             \\
7.7  & $\cdots$ & 22     &22 &20& 55  &35  &55& C-C vibration        \\
8.6  & $\cdots$ & 6     &6  &10& 35  &20  &35& C-H in-plane bend    \\
11.3 & $\cdots$ & 4      &6  &4& 36  &250  &52& C-H solo out-of-plane bend \\
11.9 & $\cdots$ & 7    &7  &7& 12  &60  &12& C-H duo out-of-plane bend  \\
12.7 & $\cdots$ & 3.5    &5&3.5& 28  &150  &28& C-H trio out-of-plane bend \\
13.6 & $\cdots$ & 4    &4  &4& 3.7 &3.7 &3.7& C-H quarto out-of-plane  \\
14.3 & $\cdots$ & 5    &5  &5& 0.9 &0.9 &0.9& C-C skeleton    vibration      \\
15.1 & 15.4 & 3    &4  &4& 0.3 &0.3 &0.3& C-C skeleton    vibration   \\
15.7 & 15.8 & 2    &5  &4& 0.3 &0.3 &5& C-C skeleton     vibration  \\
16.5 & 16.4 & 3    &10 &1& 0.5 &5   &6& C-C skeleton     vibration  \\
18.2 & 17.4 & 3    &10 &0.2& 1   &3   &1& C-C skeleton     vibration  \\
21.1 & 18.9 & 3    &10 &0.2& 2   &2   &2& C-C skeleton     vibration  \\
23.1 & $\cdots$ & 3    &10 &4& 2   &2   &1& C-C skeleton     vibration  \\
\hline
\end{tabular}
\end{center}
{\bf {Notes.}}$^a$Cross sections are integrated over the band and are
given per H atom for C-H vibrations, and per C atom for C-C
modes; $^b$Assignment following Tielens (2008), Moutou et al. (1996),
Pauzat et al. (1997); $^c$ Siebenmorgen et al. (2001); $^d$ {\it {this
    work}} (Sect.~\ref{ism.sec}, Sect.~\ref{hd37903}).
\end{table*}

\subsection {Cross sections of PAHs\label{Cpah.sec}}

Strong infrared emission bands in the 3 -- 13\,$\mu$m range are
observed in a variety of objects. The observations can be explained by
postulating as band carriers UV-pumped large PAH molecules that show
IR fluorescence. PAHs are an ubiquitous and important component of the
ISM. A recent summary of the vast literature of astronomical PAHs is
given by Joblin \& Tielens (2011).

Unfortunately, the absorption cross section of PAHs, $C_{\rm {PAH}}$,
remains uncertain. PAH cross sections vary by large factors from one
molecule species to the next and they depend strongly on their
hydrogenation coverage and charge state. Still in $C_{\rm {PAH}}$ one
notices a spectral trend: molecules have a cut-off at low (optical)
frequencies that depends on the PAH ionisation degree (Schutte et al.\
1993, Salama et al.\ 1996), a local maximum near the 2175\,\AA \,
extinction bump (Verstraete \& L\'eger\ 1992; Mulas et al.\ 2011), and a
steep rise in the far UV (Malloci et al.\ 2011; Zonca et al.\
2011).

We guide our estimates of the absorption cross section at photon
energies between 1.7--15\,eV of a mixture of ionised and neutral PAH
species to the theoretical studies by Malloci et al. (2007). We follow
Salama et al. (1996) for the cut-off frequency $\nu_{\rm {cut}}$ or
cut-off wavelength: $\lambda_{\rm {cut}}^{-1} \ = \ 1 + 3.8 \times
({0.4 \, N_{\rm C}})^{-0.5}$\,($\mu$m)$^{-1}$ and set $ \lambda_{\rm
  {cut}} \geq 0.55\mu$m. The cross section towards the near IR is
given by Mattioda et al. (2008, their Eq.~1):

\begin{equation}
  C_{\rm M}(\nu)  =
  N_{\rm C}   \  \kappa_{\rm{UV}} \ {10 ^{-1.45 \lambda}} 
  \quad {\rm {:}} \quad \nu \leq  \nu_{\rm {cut}}\,,
\label{cpahcon.eq}
\end{equation}

\noindent
where $N_{\rm {C}}$ is the number of carbon atoms of the PAHs,
$\kappa_{\rm{UV}} = 1.76 \times 10^{-19}$\, (cm$^2$/C-atom);
energetically unimportant features near 1$\mu$m are excluded. The
2175\,\AA\ bump is approximated by a Lorentzian profile
(Eq.\ref{cpahl.eq}). In the far UV at $\lambda^{-1} > 6 \mu$m$^{-1}$,
the cross section is assumed to follow that of similar sized graphite
grains. The influence of hard radiation components on PAHs is
discussed by Siebenmorgen \& Kr\"ugel (2010). The size of a PAH, which
are generally non-spherical molecules, can be estimated following
Tielens (2005), by considering the radius of a disk of a centrally
condensed compact PAH that is given by $r \sim 0.9 {N_{\rm C}}^{0.5} $
(\AA).

With the advent of {\it{ISO}} and {\it{Spitzer}} more PAH emission
features and more details of their band structures have been detected
(Tielens\ 2008).  We consider 17 emission bands and apply, for
simplicity, a damped oscillator model.  Anharmonic band shapes are not
considered despite having been observed (Peeters, et al. 2002, van
Diedenhoven et al. 2004). The Lorentzian profiles are given by

\begin{equation}
  C_{\rm L}(\nu) \ = \ N_{\rm {C, H}} \cdot \sigma_{\rm int} \; 
   {\nu_0^2\over c} \cdot
  {\gamma \nu^2 \over \pi^2 (\nu^2-\nu_0^2)^2 + (\gamma \nu / 2)^2 }\,,
\label{cpahl.eq}
\end{equation}

\noindent
where $N_{\rm {C, H}}$ is the number of carbon or hydrogen atoms of
the PAHs in the particular vibrational mode at the central frequency
$\nu_0= c/\lambda_0$ of the band, $\sigma_{\rm{int}} = \int
{\sigma_{\lambda} \mathrm{d}\lambda}$ is the cross section of the band
integrated over wavelength, and $\gamma$ is the damping constant. PAH
parameters are calibrated by Siebenmorgen \& Kr\"ugel (2007) using
mid-IR spectra of starburst nuclei and are listed in
Table~\ref{pah.tab}.  Their procedure first solved the radiative
transfer of a dust embedded stellar cluster, which contains young and
old stellar populations. A fraction of the OB stars are in compact
clouds that determine the mid IR emission. In a second step, the model
is applied to NGC\,1808, a particular starburst, and the mid-IR
cross-sections of PAHs are varied, until a satisfactory fit to the ISO
spectrum is found (Siebenmorgen et al. 2001). Finally, the so derived
PAH cross-sections are validated by matching the SED of several well
studied galaxies.  In this latter step, the PAH cross-sections are
held constant and the luminosity, size and obscuration of the star
cluster is varied (Siebenmorgen et al. 2007).  Efstathiou \&
Siebenmorgen (2009) and other colleagues, using the starburst library,
have further confirmed the applied PAH mid--IR cross--sections. The
total PAH cross section, $C_{\rm {PAH}}$, is given as the sum of the
Lorentzians (Eq.~\ref{cpahl.eq}) and the continuum absorption
(Eq.~\ref{cpahcon.eq}).

\begin{figure}
{\hspace{-0.75cm}   \includegraphics[width=9.4cm]{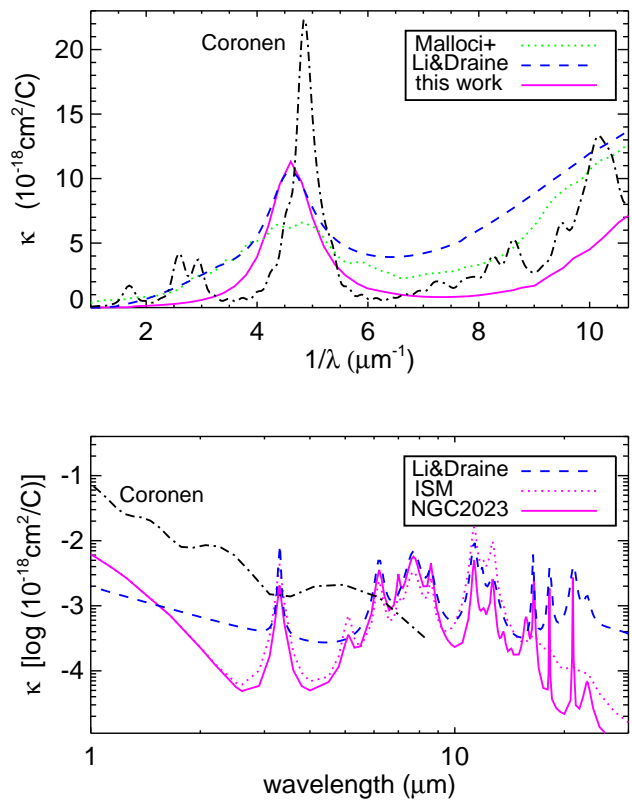}}
\caption{Absorption cross-sections of PAHs in the optical/UV (top) and
  infrared (bottom) as suggested by Li \& Draine (2001, dashed blue),
  Malloci et al. (2007, dotted green) and {\it {this work}}, with
  parameters of Table~\ref{pah.tab} for the ISM (dotted magenta) and
  for NGC\,2023 (full magenta line). For comparison we show the
  cross-section of Coronene (dash--dotted black). \label{PAHKappa.ps}}

\end{figure}

In Fig.\ref{PAHKappa.ps}, we show the PAH absorption cross-sections
suggested by Li \& Draine (2001), Malloci et al. (2007), this work, as
well as the cross-section of a particular PAH molecule,
Coronene (Mulas priv.com.).  In the far UV the PAH cross
sections agree within a factor of two. Near the 2200\,\AA \ bump we
apply for the absorption maximum the same frequency and strength as Li
\& Draine (2001) but a slightly smaller width. Our choice of the width
is guided by Coronene, which has a feature shifted to 2066\,\AA
.  Malloci et al. (2007) derive a mean PAH absorption cross section by
averaging over more than 50 individual PAHs, which are computed in
four charge states of the molecules. Such a procedure may cause a
slight overestimate of the width of the PAH band near the 2200\,\AA
\ bump, because different molecules show a peak at different
frequencies in that region (for an example, see Coronene in
Fig.\ref{PAHKappa.ps}). Between 3 and 4\,$\mu$m$^{-1}$ the cross
section by Malloci et al. (2007) and Li\& Draine (2001) are identical
and agree within a factor $\sim 3$ to that one derived for
Coronene. In the lower panel of Fig.\ref{PAHKappa.ps} PAH
absorption cross sections between $1 - 30\,\mu$m are displayed for
Coronene, neutral PAH-graphite particles with 60 C atoms (Li
\& Draine, 2001), and fitting results of this work, one for the solar
neighbourhood (labelled ISM) and a second for the reflection nebulae
NGC\,2023 (Table~\ref{pah.tab}). In the near IR, below 3\,$\mu$m, the
PAH cross sections are orders of magnitude smaller than in the
optical. The scatter in the near IR cross sections of PAH is
energetically not important for computing the emission spectrum at
longer wavelengths. In the emission bands we find similar cross
sections as Li \& Draine for neutral species whereas those for ionised
PAHs are larger by a factor of 10. Our model differs to the one by Li
\& Draine as we explain most of the observed ratios of PAH emission
bands by de-hydrogenation rather than by variations of the PAH charge
state (see Fig.1 in Siebenmorgen \& Heymann, 2012). Beyond 15$\mu$m a
continuum term is often added to the PAH cross sections (D\'{e}sert et
al. 1990; Schutte et al. 1993; Li \& Draine 2001; Siebenmorgen et
al. 2001). We neglect such a component as it requires an additional
parameter and is not important for this work.

\subsection{Dust populations \label{dustpop.sec}}
We consider two dust materials: amorphous silicates and carbon.  Dust
particles of various sizes are needed to fit an extinction curve from
the infrared to the UV. Our size range starts from the molecular domain
($r_{-} = 5$ \AA ) to an upper size limit of $r_{+} \la 0.5\,\mu$m that
we constrain by fitting the polarisation spectrum. For simplicity we
apply a power law size distribution $\mathrm{d}n(r)/\mathrm{d}r
\propto r^{-q}$ (Mathis et al. 1977).

We aim to model the linear and circular polarisation spectrum of
starlight so that some particles need to be of non--spherical shape
and partly aligned. Large homogeneous {\it spheroids} are made up of
silicate with optical constants provided by Draine (2003) and
amorphous carbon with optical constants by Zubko et al. (1996), using
their mixture labelled ACH2. The various cross sections of spheroids
are computed with the procedure outlined in Sect.~(\ref{C.sec}) for
100 particle sizes between {$60\,{\rm{\AA}} \leq r \leq 800$\,nm}. In
addition there is a population of {\it small} silicates and graphite.
For the latter we use optical constants provided by Draine (2003). The
small particles are of spherical shape and have sizes between 5\,\AA
\, $< r \la 40$\,\AA \,. We take the same exponent $q$ of the size
distribution for small and large grains.  For graphite the dielectric
function is anisotropic and average extinction efficiencies are
computed by setting ${Q} = 2 {Q}(\epsilon^\bot)/3 + {Q}(\epsilon^\Vert)/3
$, where $\epsilon^\bot$, $\epsilon^\Vert$ are dielectric constants
for two orientations of the $\vec{E}$ vector relative to the basal
plane of graphite (Draine \& Malhotra 1993). Efficiencies in both
directions are computed by Mie theory.  We include small and large
PAHs with 50 C and 20 H atoms and 250 C and 50 H atoms, respectively.
Cross sections of PAHs are detailed in Sect.~\ref{Cpah.sec}. In
summary, we consider four different dust populations, which are labelled
in the following as large silicates (Si), amorphous carbon (aC), small
silicates (sSi), graphite (gr) and PAH.

\subsection{Extinction curve \label{extin.sec}}
The attenuation of the flux of a reddened star is described by the
dust extinction $ A(\nu) = 1.086 \ \tau_{\rm ext}(\nu)$, which is
wavelength dependent and approaches zero for long wavelengths.
Extinction curves are measured through the diffuse ISM towards
hundreds of stars, and are observed from the near IR to the UV. The
curves vary for different lines of sight. The extinction curve
provides information on the composition and size distribution of the
dust. For the $B$ and $V$ photometric bands it is customary to define
the ratio of total--to--selective extinction $R_{\rm {V}} = A_{\rm
  {V}} / (A_{\rm {B}} - A_{\rm {V}})$ that varies between $2.1 \la
R_{\rm{V}} \la 5.7$. Flat extinction curves with large values of
$R_{\rm{V}}$ are measured towards denser regions.

We fit the extinction curve, or equivalently the observed optical
depth profile, along different sight lines by the extinction cross
section of the dust model, so that:

\begin{equation}
  \left({\tau (\nu) \over \tau_{\rm v}}\right)_{\rm {obs}}  \sim \left({K_{\rm {ext}}(\nu) \over K_{\rm {ext, V}}}\right)_{\rm {model}}\,,
\label{tauK.eq}
\end{equation}

\noindent
where $K_{\rm ext}({\nu})$ is the total mass extinction cross section
averaged over the dust size distribution in (cm$^2$/g-dust) given by

\begin{equation}
K_{\rm ext} = \sum_i \ \int_{r_-}^{r_+} K_{{\rm ext}, i}(r) \ \mathrm{d}r\,.
\label{Ktot.eq}
\end{equation}

\noindent
where index $i$ refers to the dust populations
(Sect.~\ref{dustpop.sec}).  The extinction cross sections $K_{{\rm
    {ext}}, i}(r)$\, (cm$^2$/g-dust) of a particle of population $i
\in \{ {\rm {Si, aC, sSi, gr}} \}$, of radius $r$ and density $\rho_i$
are

\begin{equation}
K_{{\rm ext}, i}(r) =    {w_i \over {{ \displaystyle 4 \pi \over 3} \ \rho_i}} \
{r^{-q} \over   \displaystyle  \int_{r_{-,i}}^{r_{+,i}} r^{3-q} \  \mathrm{d}r} \ C_{{\rm ext}, i}(r)\,,
\label{K.eq}
\end{equation}

\noindent
where $w_i$ is the relative weight of dust component $i$, which, for
large amorphous carbon grains, is

\begin{equation}
w_{\rm {aC}} = {{\Upsilon_{\rm {aC}} \ \mu_{\rm C}} \over
{({\tiny{\Upsilon_{\rm {aC}}+\Upsilon_{\rm {gr}}+\Upsilon_{\rm {PAH}}) \mu_{\rm C}
+ (\Upsilon_{\rm {Si}}+\Upsilon_{\rm {sSi}})  \mu_{\rm Si}}}}}\; ,
\label{w.eq}
\end{equation}

\noindent
with molecular weight of carbon $\mu_{\rm C} =12$ and silicate grains
$\mu_{\rm Si} =168$.  As bulk density we take $\rho_C \sim
2.3$\,(g/cm$^3$) for all carbon materials and $\rho_{\rm Si} \sim
3$\,(g/cm$^3$). Dust abundances are denoted by $\Upsilon$ together
with a subscript for each dust population (Sect.~\ref{dustpop.sec}).
The expressions of the relative weights of the other grain materials
are similar to expression Eq.~(\ref{w.eq}). The cross section
normalised per gram dust of a PAH molecule is

\begin{equation}
  K_{\nu{\rm {, PAH}}} =     {w_{_{ \rm {PAH}}} \over N_{\rm C} \ 
                          \mu_{\rm C} \ m_{\rm p} } \ C_{\nu \rm{, PAH}}\,,
\label{Kpah.eq}
\end{equation}

\noindent
where $m_{\rm p}$ is the proton mass.

\begin{figure}
\hspace{-1cm}   
\includegraphics[width=10cm]{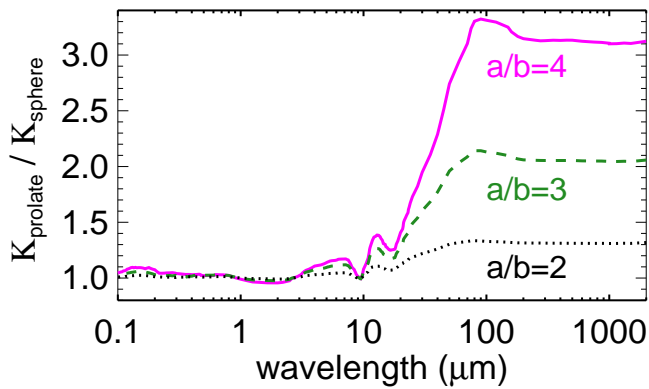}
\caption{Ratio of the total mass extinction cross section of large
  prolate and spherical particles with same volume. Prolates, with
  $a/b$ ratios as labelled, and spheres have same size distribution
  and relative weights ($w_i$) as for dust in the solar neighbourhood
  (Table~\ref{para.tab}).  \label{cmpC.fig} }
\end{figure}

Our grain model with two types of bare material is certainly
simplistic. ISM dust grains are bombarded by cosmic rays and atoms,
they grow and get sputtered. Therefore fluffy structures with
impurities and irregular grain shapes are more realistic. The cross
section of composite particles (Kr\"ugel \& Siebenmorgen 1994,
Ossenkopf \& Henning 1994, Voshchinnikov et al. 2005), that are porous
aggregates made up of silicate and carbon, vary when compared to
homogeneous particles by a factor of 2 in the optical and, and by
larger factors in the far IR/submm. We study the influence of the
grain geometry on the cross section. For this we compare the cross
section of large prolate particles $K_{\rm {prolate}}$ with axial
ratios $a/b = 2$, 3, and 4, to that of spherical grains $K_{\rm
  {sphere}}$ (Fig.~\ref{cmpC.fig}). With the exception of
Fig.~\ref{cmpC.fig}, throughout this work for large grains we use the
cross-sections computed for spheroids.  The dust models with the two
distinct grain shapes are treated with same size and mass distribution
as the large ISM grains above. The peak-to-peak variation of the ratio
$K_{\rm {prolate}} / K_{\rm {sphere}}$ is for $\lambda \leq 2\mu$m:
$\sim 4$\% for an axial ratio of $a/b=2$, 9\% for $a/b=3$ and 14\% for
$a/b=4$, respectively. In the far IR the prolate particles have by a
factor of 1.5 -- 3 larger cross sections than spherical grains.  In
that wavelength region the cross section varies roughly as $C_{abs}
\propto \nu^2$, so that the emissivity of a grain with radius $r$ is
about $\epsilon \propto r \ T_{\rm d}^6$, where $T_{\rm d}$ denotes
the dust temperature. Therefore spheroids with $a/b \leq 2$ obtain in
the same radiation environment $\sim 10\%$ lower temperatures than
their spherical cousins, an effect becoming larger for more elongated
particles (Voshchinnikov et al., 1999). Nevertheless, the larger far
IR/submm cross sections of the spheroids scales with the derived mass
estimates of the cloud ($M \propto 1/K$, in the optical thin case) and
is therefore important.

We keep in mind the above mentioned simplifications and uncertainties
of the absorption and scattering cross sections $K_i$ and allow for
some fine-tuning of them, so that the observed extinction curve,
$K^{\rm {obs}}_{\rm {ext}}(\nu) = K_{\rm {V}} \left( \tau({\nu}) /
\tau_{\rm V}\right)_{\rm {obs}}$ (Eq.~\ref{tauK.eq}) is perfectly
matched.  Another possibility to arrive at a perfect match of the
extinction curve can be achieved by ignoring uncertainties in the
cross sections and altering the dust size distribution (Kim et
al. 1994, Weingartner \& Draine 2001, Zubko et al. 2004).  In our
procedure we apply initially the $K_i$'s as computed strictly
following the prescription of Sect.(\ref{C.sec}) and
Eq.~(\ref{K.eq}). Then for each wavelength new absorption and
scattering cross sections $K_i' = f \ K_i$ are derived using

\begin{equation}
f  =  {K^{\rm {obs}}_{\rm {ext}} \over K_{\rm {sca}}} \ \Lambda\,,
\label{f.eq}
\end{equation}

\noindent
where $0 < \Lambda \leq 1$ denotes the dust albedo from the initial
(unscaled) cross sections
\begin{equation}
\Lambda = K_{\rm {sca}} / K_{\rm {ext}}\,.
\label{albedo.eq}
\end{equation}

\noindent
We note that the procedure of Eq.~(\ref{f.eq}) is only a fine
adjustment. We vary the $K_i$'s of the large spheroids only at
$\lambda < 2\mu$m and the $K_i$'s of the small grains only in the UV
at $\lambda < 0.3\mu$m. At these wavelengths we allow variations of
the cross sections by never more than 10\%, so we set $\min {(f)} >
0.9$ and $\max {(f)} < 1.1$. Typically a few \% variation of the
initial cross sections is sufficient to perfectly match the observed
extinction profile; otherwise cross sections remain unchanged.  In
Fig.~\ref{ism.fig} {\it {(top)}} both extinction models are indicated,
the one derived from the unscaled cross sections is labelled ``fit''
and the other, using scaled cross sections, is marked ``best''. The
uncertainty of our physical description in explaining the extinction
curve is measured by the $f$--value. From this we conclude that the
initial model is accurate down to a few \%, which is within the
observational uncertainties. Therefore we prefer keeping the
description of the dust size distribution simple.

\subsection {Element abundances \label{abu.sec}}

Estimating the absolute elemental abundances is a tricky task, and in
the literature there is no consensus yet as to their precise values.
To give one example, for the cosmic (solar or stellar) C/H abundance
ratio, expressed in ppm, one finds values of: 417 (Cameron \&
Fegley\ 1982), 363 (Anders \& Grevesse\ 1989), 398 (Grevesse et
al.\ 1993), 330 (Grevesse \& Sauval\ 1998), 391 (Holweger\ 2001), 245
(Asplund \& Garcia-Perez\ 2004), 269 (Asplund et al.\ 2009), 316
(Caffau et al.\ 2010), 245 (Lodders\ 2010), and 214 (Nieva \&
Przybilla\ 2012), respectively.  Towards 21 sight lines Parvathi et
al. (2012) derive C/H ratios between 69 and 414\,ppm.  Nozawa \&
Fukugita (2013) consider a solar abundance of C/H = 251\,ppm with a
scatter between 125 to 500\,ppm.  For dust models one extra
complication appears as one needs to estimate how much of the carbon
is depleted from the gas into the grains. Present estimates are that
60\% -- 70\% of all C atoms stick into dust particles (Sofia et
al. 2011, Cardelli 1996), whereas earlier values range between 30 --
40 \% (Sofia et al. 2004). From extinction fitting, Mulas et
al. (2013) derive an average C abundance in grains of $145$\,ppm and
estimate an uncertainty of about a factor of two. The abundance of O
is uncertain within a factor of two. Variations of absolute abundance
estimates are noticed for elements such as Si, Mg and Fe, for which
one assumes that they are completely condensed (for a recent review on
dust abundances see Voshchinnikov et al. 2012). Averaging over all
stars of the Voshchinnikov \& Henning (2010) sample the total Si
abundance is $25 \pm 3$\,ppm.

We design a dust model where only relative abundances need to be
specified. These are the weight factors $w_i$ as introduced in
Eq.~(\ref{w.eq}). The weight factors prevent us from introducing
systematic errors of the absolute dust abundances into the model.
Still they can be easily converted into absolute abundances of element
$i$ in the dust.  We find for the solar neighbourhood a total C
abundance of $w_{\rm {C}} = 37.2$\% (Table ~\ref{para.tab}). To
exemplify matters let us assume that the absolute C abundance in dust
is $183$\,ppm and of Si of $22$\,ppm. For this case one converts the
weight factors into absolute element abundances of the dust
populations to $\Upsilon_{\rm {aC}} = 143.5$\,ppm, $\Upsilon_{\rm
  {Si}} = 19.1$\,ppm, $\Upsilon_{\rm {gr}} = 21$\,ppm, $\Upsilon_{\rm
  {sSi}} = 2.9$\,ppm, $\Upsilon_{\rm {PAHs}} = 6.7$\,ppm,
$\Upsilon_{\rm {PAHb}} = 11.5$\,ppm, respectively. These numbers can
be updated following Eq.~\ref{w.eq} and Table~\ref{para.tab} whenever
more accurate estimates of absolute element abundances in the dust
become available.


\subsection {Optical thin emission}

For optically thin regions we model the emission spectrum of the
source computed for 1g of dust at a given temperature.  The emission
$\epsilon_i (r)$ of a dust particle of material $i$ and radius $r$ is

\begin{equation}
\begin{array} {r l}
\epsilon_{i}(r) =&  \mathlarger{\int} {K_{\nu{{, i}}}^{abs} (r) \, J_{\nu} \, \mathrm{d}\nu} \\
 & \\
=& \mathlarger{\int} {K_{\nu{{, i}}}^{abs} (r) \, P(r,T) \, B_{\nu}(T) \, \mathrm{d}T \, \mathrm{d}\nu }\,, 
\label{emis.eq}
\end{array}
\end{equation}

\noindent
where the mass absorption cross sections are defined in
Eqs.~(\ref{K.eq},~\ref{Kpah.eq}), $J_{\nu}$ denotes the mean
intensity, $B_{\nu}(T)$ is the Planck function and $P(r,T)$ is the
temperature distribution function that gives the probability of
finding a particle of material $i$ and radius $r$ at temperature
$T$. This function is evaluated using an iterative scheme that is
described by Kr\"ugel (2008). The $P(T)$ function needs only to be
evaluated for small grains as it approaches a $\delta$-function for
large particles. The total emission $\epsilon_{\nu}$, is given as sum
of the emission $\epsilon_{{i,} \nu}(r)$ of all dust components.

\subsection {Dust radiative transfer}

For dust enshrouded sources we compute their emission spectrum by
solving the radiative transfer problem. Dropping for clarity the
frequency dependency of the variables, the radiative transfer equation
of the intensity $I$ is

\begin{equation}
  I(\tau) = I(0) \ e^{-\tau} \ + \  \int_0^{\tau}  S(\tau') \ e^{-(\tau - \tau')}  \mathrm{d}\tau'\,.
\end{equation}

\noindent
We take as  source function
\begin{equation}
S  =  {{K_{\rm {sca}} J  + \sum_{i} \epsilon_i} \over {K_{\rm {ext}}}}\,,
\end{equation}
where $\epsilon_i$ is the emission of dust component $i$ computed
according to Eq.~(\ref{emis.eq}). The problem is solved by ray tracing
and with the code described in Kr\"ugel (2008). Dust temperatures and
$P(T)$ are derived at various distances from the source. The star is
placed at the centre of the cloud, and is considered to be spherically
symmetric. Our solution of the problem for arbitrary dust geometries
is discussed by Heymann \& Siebenmorgen (2012).


\begin{figure}
\hspace{-0.7cm}
   \includegraphics[width=10.cm]{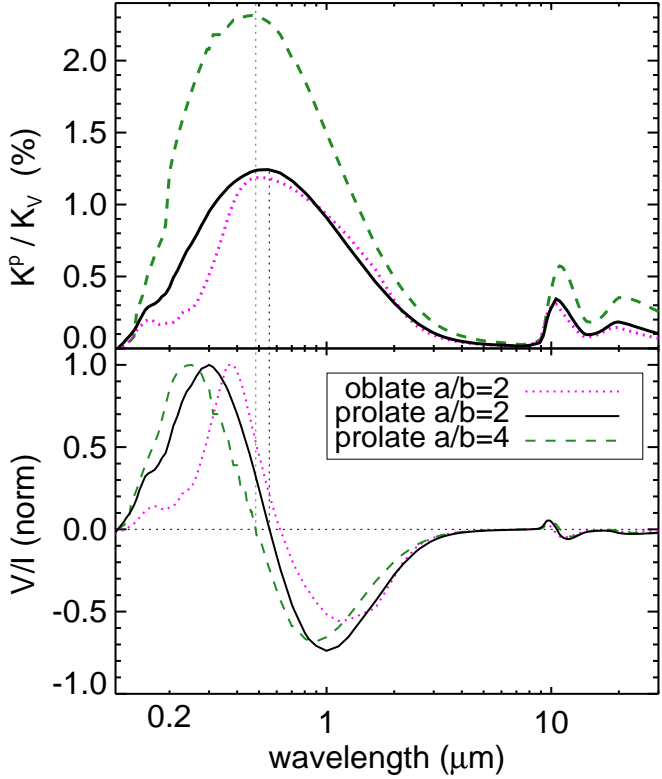}
   \caption{Linear (top) and circular (bottom) polarisation spectra of
     silicates with $r_{-}=100$\,nm, $r_{+}=450$\,nm, $q=3.5$. The axial
     ratios $a/b$ and grain shapes (prolate, oblate) are
     indicated. The circular polarisation spectra are normalized to
     their maxima.\label{Polab.fig} }
\end{figure}


\begin{figure}
   \includegraphics[width=8.cm]{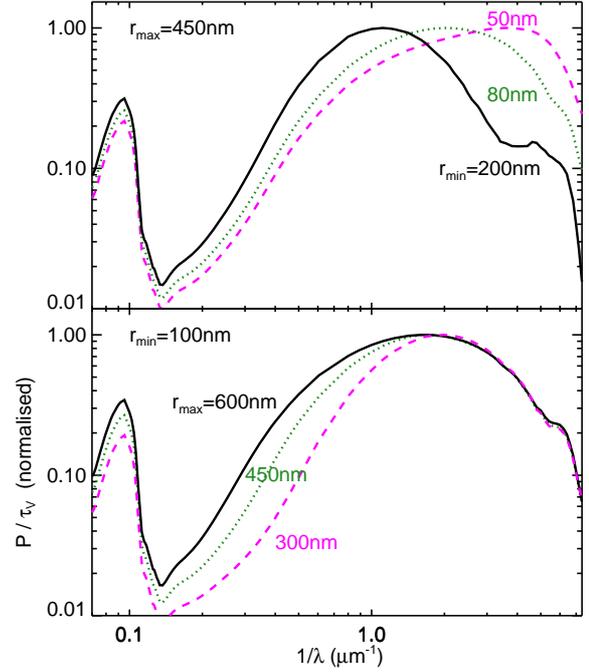}
   \caption{Influence of the upper, $r_{+}$, and lower, $r_{-}$, limit
     of the particle radii of aligned silicates on the spectral shape
     of the dichroic polarisation.  Shown are prolates with $a/b=2$
     and $q=3.5$. Top: $r_{+}$ is held constant at 450\,nm
     and polarisation spectra are computed for $r_{-} = 50$, 80, and
     200\,nm that give rise to a maximum polarisation of $p/\tau_{\rm
       V}=2.1$\%, 1.6\%, and 0.8\%, respectively. Bottom: 
     lower limit of $r_{-} = 100$\,nm and
     varying $r_{+} = 300$, 450, and 600\,nm that produce a maximum
     polarisation of $p/\tau_{\rm V}=1.7$\%, 1.4\%, and 1.2\%,
     respectively. \label{PolRpmax.fig}}
\end{figure}

\begin{figure}
   \includegraphics[width=8.cm]{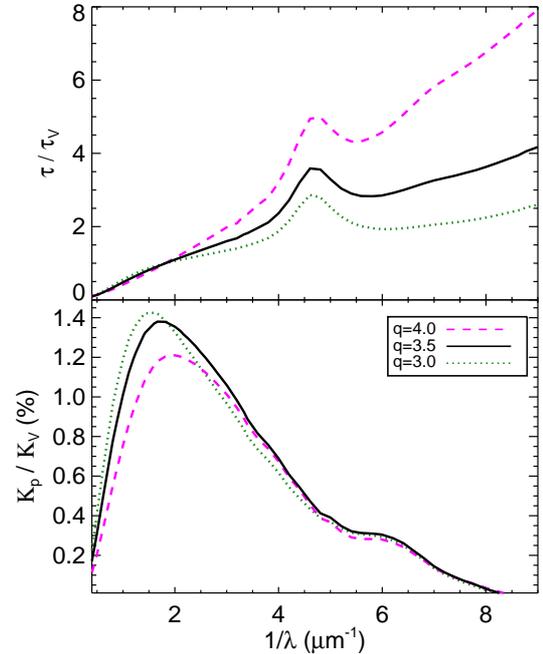}
   \caption{Influence of the exponent of the dust size distribution
     $q$ on the extinction and linear polarisation curve.  For the
     polarisation we consider prolates made up of silicates with
     $r_{+}=450$\,nm and other parameters as for the solar neighbourhood
     (Table~\ref{para.tab}). \label{ExtPolq.fig}}
\end{figure}

\subsection {Linear polarisation \label{ismpol.sec}}

Observations of reddened stars frequently show linear polarisation of
several percent. These stars have often such thin dust shells, if any,
that the polarisation cannot be explained by circumstellar dust
(Scicluna et al. 2013). As discussed in Sect.~\ref{spheroids.sec},
partly aligned non--spherical grains have different extinction with
respect to their orientation, and therefore they polarise the
radiation. The polarisation scales with the amount of dust, hence with
the optical depth towards the particular sight line
(Eqs.~\ref{tauCext.eq}, \ref{pCp.eq}).  In the model the linear
polarisation cross section $K_{\rm {p}}({\nu})$, is computed utilizing
${C}_{{\rm p}}(\nu)$ (Eq.~\ref{Cp.eq}) and replacing subscript {\it
  {ext}} by {\it p} in Eqs.~(\ref{Ktot.eq}, \ref{K.eq}). Observations
of linear polarisation by dichroic extinction are modelled using:

\begin{equation}
p/\tau_{\rm V} = K_{\rm {p}}(\nu) / K_{\rm {ext, V}}\,.
\label{ptau.eq}
\end{equation}

\noindent
We apply the dust model with the parameters of Table~\ref{para.tab},
col.2. For grain alignment we assume the IDG mechanism of
Eq. (\ref{align}). We consider moderately elongated particles with
$a/b=2$ and assume that only large grains with sufficient inertia are
aligned and that grains smaller than about 50\,nm are randomly
oriented. We show in Fig.~\ref{Polab.fig} that larger axial ratios
increase the maximum polarisation and do not influence the spectral
shape of the polarisation curve strongly.  However, for wavelengths
below $\lambda_{\rm {max}}$, one notices that oblate particles have a
stronger decline in the polarisation than prolates.

\begin{table*}
\caption{Log of FORS spectro-polarimetric observations, all of which were obtained on
2011-01-21.\label{Obs.tab} }
\begin{center}
\begin{tabular}{|ll|r|c|r|c|r|c|r|c|r|}
\hline \hline
\multicolumn{3}{|c} {Target} &      
\multicolumn{4}{|c|}{Linear polarisation} &
\multicolumn{4}{c|}{Circular polarisation} \\
\hline
  \multicolumn{2}{|c|} {Name}     & PA &
\multicolumn{2}{c|}{grism 600\,B}&
\multicolumn{2}{c|}{grism 1200\,R}&
\multicolumn{2}{c|}{grism 600\,B}&
\multicolumn{2}{c|}{grism 1200\,R}\\
\hline
\multicolumn{2}{|c|}{ } &(\degr)&
\multicolumn{1}{c|}{(hh:mm)}      &
\multicolumn{1}{c|}{$t$ (sec)}        &
\multicolumn{1}{c|}{(hh:mm)}      &
\multicolumn{1}{c|}{$t$ (sec)}        &
\multicolumn{1}{c|}{(hh:mm)}      &
\multicolumn{1}{c|}{$t$ (sec)}        &
\multicolumn{1}{c|}{(hh:mm)}      &
\multicolumn{1}{c|}{$t$ (sec)}       \\
\hline
HD\,35149 &=\,HR~1770      &  0  & 00:24 &  29 & 00:44 &  24 & 00:49 &  34 & 00:52 &  40 \\
         &             & 90  & 01:38 &  12 & 01:21 &  16 & 01:47 &  20 & 01:29 &  32 \\[2mm]
HD\,37061 &=\,$\nu$\,Ori   & 30  & 02:04 &  40 & 02:23 &  40 & 02:12 &  85 & 02:32 & 101 \\
         &             & 60  & 04:10 &  60 &   $\cdots$ &  $\cdots$& 04:17 &  60 & $\cdots$   & $\cdots$ \\
         &             & 90  & 03:51 &  90 &   $\cdots$ &  $\cdots$& 04:58 &  60 & $\cdots$   &  $\cdots$ \\
         &             &120  & 03:08 &  51 & 03:27 &  36 & 04:18 & 150 & 00:00 & 300 \\
         &             &150  & 04:28 &  56 & 00:00 & 300 & 04:35 &  60 &  $\cdots$  &  $\cdots$\\
         &             &210  & 02:48 &  12 & 00:00 & 300 & 02:56 &  48 &  $\cdots$  &  $\cdots$\\[2mm]
HD\,37903 &=\,BD$-$02\,1345&  0  & 04:49 & 150 & 05:12 & 175 & 05:00 & 280 & 05:25 & 435 \\
         &             &  0  & 05:54 & 200 & 06:12 & 235 & 06:03 & 200 & 06:22 & 240 \\[2mm]
HD\,93250 &=\,CD$-$58\,3537&  0  & 06:47 &  67 & 07:07 &  80 & 06:59 & 160 & 07:19 & 115 \\ 
         &             & 90  & 07:36 &  25 & 07:57 & 140 & 07:46 & 200 & 08:05 & 180 \\[2mm]
HD\,99872 &=\,HR~4425      &  0  & 09:16 &  40 & 09:37 &  80 & 09:22 &  40 & 09:30 &  70 \\[2mm] 
HD\,94660 &=\,HR~4263      &  0  & 08:27 &  24 & 00:00 & 300 & 08:34 &  28 &  $\cdots$  &  $\cdots$\\[2mm] 
Ve 6-23  &=\,Hen\,3-248   &  0  & 09:00 & 720 &  $\cdots$  &  $\cdots$&   $\cdots$ & $\cdots$ &  $\cdots$  &  $\cdots$     \\
\hline
\end{tabular}
\end{center}
\end{table*}


One can see from Fig.~\ref{PolRpmax.fig}, that the choice of the
lower, $r_{\rm {-}}$, and the upper particle radius $r_{+}$ of aligned
grains is sensitive to the curvature of the derived polarisation
spectrum.  Reducing the lower size limit of aligned grains, $r_{-}$,
enhances the polarisation at short wavelengths and increasing upper
size limit, $r_{+}$, produces stronger polarisation at longer
wavelengths.  A similar trend is given by altering the exponent $q$ of
the size distribution. For larger $q$ there are smaller particles
and the polarisation shifts to shorter wavelengths, the maximum
polarisation shrinks and the spectrum broadens. The increase of
$r_{\rm {-}}$ and the decrease of $q$ can be associated with the
growth of dust grains due to accretion and coagulation
processes. Voshchinnikov et al. (2013) find that both mechanisms shift
the maximum polarisation to longer wavelengths and narrow the
polarisation curve. These characteristics are parameterised by the coefficients $\lambda_{\max}$ and $k_p$ of
the Serkowski curve Eq.~(\ref{serk.eq}), respectively. However, this
effect is less pronounced than altering the range of particle sizes of
aligned grains. In Fig.\ref{ExtPolq.fig} we show the polarisation
curve for $q = 3$, 3.5, and 4 of silicates with prolate shape $a/b=2$,
$r_{\rm {-}} = 100$\,nm, $r_{+}=450$\,nm using IDG alignment as well as
the influence of $q$ on the derived extinction curve.  One notices a
strong effect in which larger $q$ values produce steeper UV
extinction.  In summary the extinction is sensitive to variations of
$q$ while the polarisation spectrum depends critically on the size
spectrum of aligned grains.

The model predicts a strong polarisation in the silicate band
(Fig.~\ref{PolRpmax.fig}). This is in agreement with the many
detections of a polarised signal at that wavelength (Smith et
al. 2000). However, in the mid IR two orthogonal mechanisms may be at
work and produce the observed linear polarisation. There is either
dichroic absorption as discussed in this work and observed in
proto-stellar systems (Siebenmorgen \& Kr\"ugel 2000) or dichroic
emission by elongated dust particles as observed on galactic scales
(Siebenmorgen et al., 2001).

The angle between the line of sight and the (unsigned) magnetic field is in the
limits $0\degr \leq \Omega \leq 90\degr$. We find that the spectral
shape of the linear polarisation only marginally depends on
$\Omega$, contrary to the maximum of the linear polarisation.  The
polarisation is strongest for $\Omega =90\degr$. For prolate silicate
particles with size distribution and IDG alignment characteristic of
the ISM the maximum polarisation decreases for decreasing
$\Omega$. For $\Omega = 60^0$ the polarisation decreases to 60\% of
that found at maximum, and for $\Omega=30\degr$ further down to
$\sim$\,20\%. The dependency of $p$ with $\Omega$ is used by
Voshchinnikov (2012) to estimate the orientation of the magnetic field
in the direction of the polarised source. Unless otherwise stated we
use $\Omega=90^{\rm {o}}$.

\subsection {Circular polarisation}

The dust model also predicts the observed circular polarisation of
light (Martin \& Campbell 1976, Martin 1978). The circular
polarisation spectrum as of Eq. (\ref{cp.eq}) is shown in
Fig.~\ref{Polab.fig}, normalised to the maximum of $V/I$. We apply the
same dust parameters as for the linear polarisation spectrum described
above. We note that $V/I$ changes sign at wavelengths close to the
position of maximum of the linear polarisation (Voshchinnikov
2004). This may explain many null detections of circular polarisation
in the visual part of the spectrum because $\lambda_{\rm {max}} \sim
0.5\mu$m. The local maxima and minima of $V/I$ critically depend on
elongation and geometry of the grain: prolate versus oblate.  In fact,
that circular polarisation can provide new insights into the optical
anisotropy of the ISM has been proposed long time ago (van de Hulst
1957, Kemp \& Wolstencroft 1972). Typically, for both particle shapes,
the maximum of $V/I$ is $\sim 7 \times 10^{-5}$ for $a/b=2$, and $\sim
25 \times 10^{-5}$ for $a/b=4$. Detecting this amount polarisation is
at the very limit of the observational capabilities of the current
instrumentation.

\section{Spectro-polarimetric observations} \label{obs.sec}

Linear and circular spectro-polarimetric observations were obtained
with the FORS instrument of the VLT (Appenzeller et al. 1998).  Our
main goal is to constrain the dust models with ultra-high accuracy
polarisation measurements. Stars were selected from the sample
provided by Voshchinnikov \& Henning (2010). Towards these sight lines
linear polarisation was previously detected and extinction curves are
available. Targets were chosen based on visibility constraints at the time of
the observations. Among the various available grisms, we adopted those
with higher resolution.  The observed wavelength ranges are
340-610\,nm in grism 600\,B, and 580-730\,nm in grism 1200\,R. This
grism choice was determined by practical considerations on how to accumulate
a very high signal-to-noise ratio (SNR) in the interval range where we
expect a change of sign of the circular
polarisation. Table~\ref{Obs.tab} gives our target list, the
instrument position angle on sky (counted counterclockwise from North
to East), the UT at mid exposure, and the total exposure time in
seconds for each setting.

Linear polarimetric measurements were obtained by setting the
$\lambda/2$ retarder waveplate at position angles 0\degr, 22\fdg5,
45\degr, and 67\fdg5. All circular polarisation measurements were
obtained by executing once or twice the sequence with the $\lambda/4$
retarder waveplate at 315\degr, 45\degr, 135\degr, and
225\degr. Because targets are bright, to minimise risk of saturation
we set the slit width to $0\farcs5$; this provides a spectral
resolution of $\sim 1500$ and $\sim 4200$ in grism 600\,B and 1200\,R,
respectively. Observations of HD\,99872 and Ve\,6-23 were performed
with a slit width of 1\arcsec\, providing a spectral resolution of
$\sim 800$ and $\sim 1200$ in grism 600\,B and 1200\,R, respectively.
Finally, spectra are rebinned by 256 pixels to achieve the highest
possible precision in the continuum. This gives a spectral bin of
$\sim 17$\,nm and $\sim 10$\,nm in grisms 600\,B and 1200\,R,
respectively. It allows us to push the signal-to-noise ratio of the
circular polarisation measurements to a level of several tens of
thousands over a 10\,nm spectral bin.
\subsection{Standard stars}\label{Sect_Standard}

To verify the alignment of the polarimetric optics, and to measure the
instrumental polarisation, we observed two standard stars: HD\,94660
and Ve\,6-23. From the circular polarimetric observations of the
magnetic star HD\,94660 we expected a zero signal in the continuum,
and, in spectral lines, a signal consistent with a mean longitudinal
magnetic field of about -2\,kG (Bagnulo et al. 2002).  Indeed we found
in the continuum a circular polarisation signal consistent with zero.
The magnetic field was measured following the technique described in
Bagnulo et al. (2002) on non-rebinned data. We found a value
consistent with the one above and concluded that the $\lambda/4$
retarder waveplate was correctly aligned.

In Ve\,6-23 we measured a linear polarisation consistent with
the expected values of about 7.1\,\% in the $B$ band, 7.9\,\% in the
$V$ band, and with a position angle of $\sim 173\degr$\ and 172\degr\ in
$B$ and $V$, respectively (Fossati et al. 2007). This demonstrates that
the $\lambda/2$ retarder waveplate was correctly set.  An unexpected
variation of the position angle $\Theta$ is observed at $\lambda \la
400$\,nm, possibly related to a substantial drop in the
signal-to-noise ratio. Patat \& Romaniello (2006) identified such a
spurious and asymmetric polarization field that is visible in FORS1
imaging through the B band. We measured the linear polarisation of
HD\,94660, expecting a very low level in the continuum, and we found a
signal of $\sim 0.18$\,\% that is discussed below.


\begin{figure}
\includegraphics[width=9.cm]{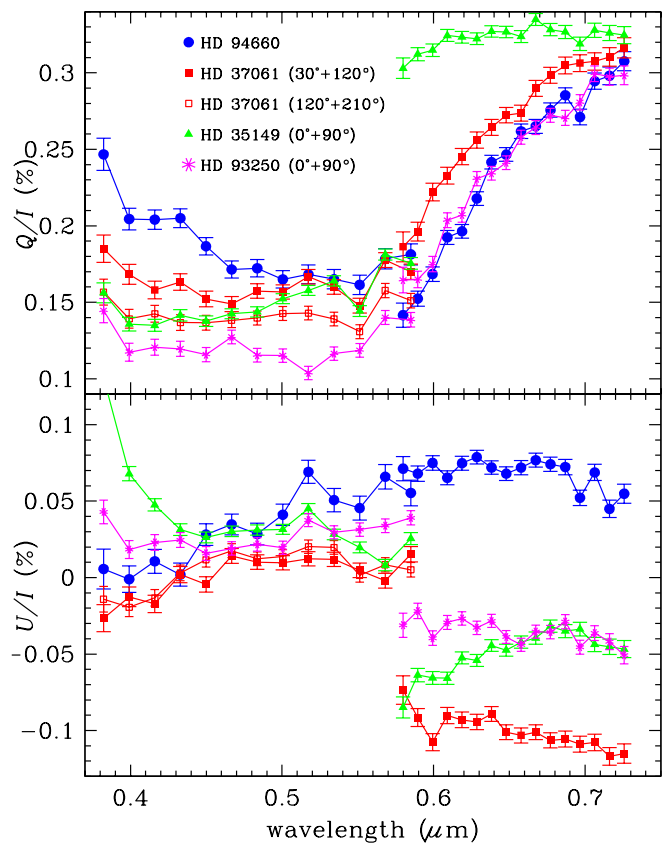}
\caption{\label{Fig_LIN_IP} Instrumental linear polarisation spectra
  derived from observing pairs as labelled together with the continuum
  polarisation of HD\,94660.  Stokes parameters $Q/I$ (top) and
  $U/I$ (bottom) are measured with respect to the instrument
  reference system.}

\end{figure}


\begin{figure}
\includegraphics[width=9.cm]{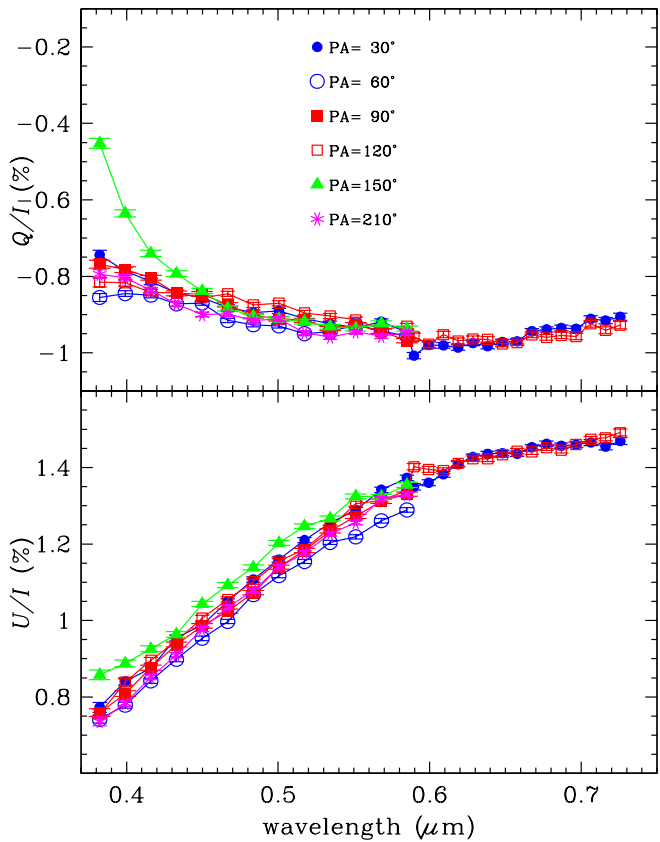}
\caption{\label{Fig_HD37061} Linear polarisation spectra of HD\,37061
  that are corrected for instrumental contribution at various PA as
  labelled.  Stokes parameters $Q/I$ (top) and $U$ (bottom)
  are measured having as a reference direction the celestial meridian
  passing through the target.  }
\end{figure}


\begin{figure}
\includegraphics[width=8.5cm]{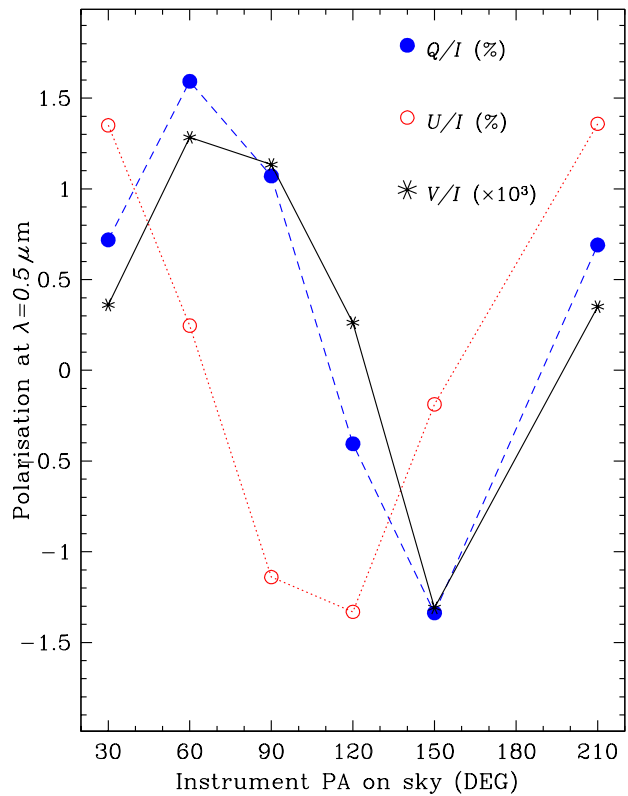}
\caption{\label{Fig_XTalk} Polarisation at $\lambda = 500$\,nm of
  HD\,37061 at various instrument position angles. $Q/I$ (blue dashed
  line) and $U/I$ (red dotted line) are measured in the instrument
  reference system.  The circular polarisation (black solid line) is
  approximately one tenth of the linear polarisation in the principal
  plane of the Wollaston prism. }
\end{figure}

\begin{figure}
\includegraphics[width=9cm]{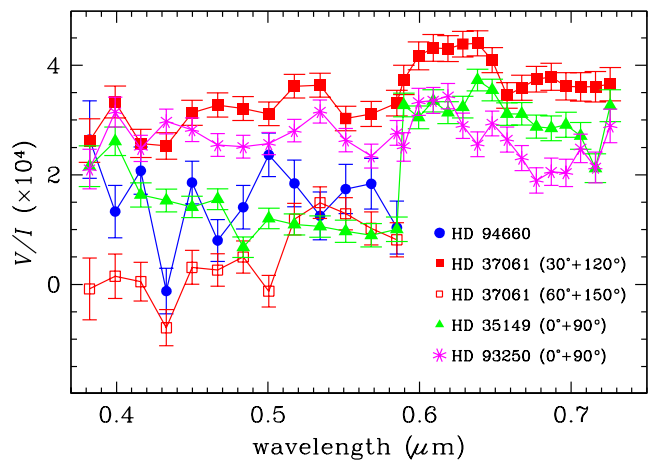}
\vspace{-5cm}
\caption{\label{Fig_CIR_IP} Circular polarisation in the continuum of
  HD\,94660 at instrument PA=$0^\circ$ and of other stars from
  observing pairs as labelled.}
\end{figure}

\subsection{Instrumental linear polarisation}\label{Sect_LIP}

We observed the stars at different instrument position angles on sky
to evaluate the instrumental polarisation. The rationale of this
observing strategy is that observations of circular polarisation
should not depend on the position angle (PA) of the instrument, while
linear polarisation measurement follow a well known transformation
(see Eq.~(10) in Bagnulo et al. 2009). For example, denoting with
$P_Q^{(\alpha)}$, $P_U^{(\alpha)}$, $P_V^{(\alpha)}$ the reduced
Stokes parameters measured with the instrument position at PA=
$\alpha$ on sky, one should find:

\begin{equation}
\label{Eq_Rot}
\begin{array}{rcl}
P_Q^{(90)} &=& -P_Q^{(0)}\\[2mm]
P_U^{(90)} &=& -P_U^{(0)}\\[2mm]
P_V^{(90)} &=&  P_V^{(0)}\; . \\
\end{array}
\end{equation}

Departures from this behaviour may be due to spurious instrumental
effects, which should not change as the instrument rotates.  Hough et
al. (2007) suggest to evaluate the instrumental linear polarisation as

\begin{equation}
\begin{array}{rcl} 
P_Q^{\rm instr} &=& \frac{1}{2}\, \left(P_Q^{(0)} - P_Q^{(90)}\right)\\
P_U^{\rm instr} &=& \frac{1}{2}\, \left(P_U^{(0)} - P_U^{(90)}\right) \, . \\
\end{array}
\end{equation}

Instrumental polarisation of $\sim 0.1$\,\% is identified in FORS1
data by Fossati et al. (2007), and is discussed by Bagnulo (2011).
From the analysis of our data we confirm that FORS2 measurements are
affected by an instrumental polarisation that depends on the adopted
grism. By combining the measurements of HD\,37061 obtained at
instrument position angles 30\degr\ and 120\degr\ in grism 600\,B, we
measure a spurious signal of linear polarisation of about $0.16$\,\%
that is nearly constant with wavelength along the principal plane of
the Wollaston prism.  In grism 1200\,R, we find that the instrumental
contribution in the principal plane of the Wollaston prism is linearly
changing with wavelength from $\sim 0.2$\,\% at $\lambda=580$\,nm to
$\sim 0.32$\,\% at $\lambda=720$\,nm; and in the perpendicular plane
we find a value $\sim -0.1$\,\% that is nearly constant with
wavelength (Fig.~\ref{Fig_LIN_IP}).  From the observations of the same
star obtained at instrument position angles 120\degr\ and 210\degr\
(grism 600\,B only) we retrieve $\sim 0.14$\,\% in the principal plane
of the Wollaston prism. Similar values of the instrumental
polarisation are derived for the observations of the other two targets
that are shown in Fig.~\ref{Fig_LIN_IP}. A higher instrumental
polarisation is observed at $\lambda \la 400$\,nm than at longer
wavelengths.

We conclude that the instrumental polarisation is either not constant
in time or depends on the telescope pointing. It is, at least in part,
related to the grism, and is much higher and wavelength dependent in
in the measurements obtained with the holographic grism 1200\,R than
in those obtained with grism 600\,B. We note that holographic grisms
are known to have a transmission that strongly depends on the
polarisation of the incoming radiation. Nevertheless, the instrumental
polarisation may depend also on the telescope optics and the position
of the Longitudinal Atmospheric Dispersion Corrector (LADC, Avila et
al. 1997). Our experiments to measure the instrumental polarisation by
observing at different PA suggest that the linear polarisation signal
measured in HD\,94660 is at most instrumental (Fig.~\ref{Fig_LIN_IP}).

Figure~\ref{Fig_HD37061} shows all our measurements of HD\,37061
obtained at various instrument PA and corrected for our estimated
instrumental polarisation. All these spectra refer to the north
celestial meridian.  These data show that, although the photon noise
in a spectral bin of 10\,nm is well below 0.01\,\%, the accuracy of
our linear polarisation measurements is limited by an instrumental
effect that we have calibrated probably to within $\sim 0.05$\,\%.

\subsection{Instrumental circular polarisation}\label{Sect_CIP}

In our science targets, we also find that circular polarisation
measurements depend on the instrument PA. The fact that a zero
circular polarisation is measured in the continuum of HD\,94660, a star
not linearly polarised, suggests that the cross-talk from intensity
$I$ to Stokes $V$ is negligible, and that the spurious circular
polarisation is due to cross-talk from linear to circular
polarisation. The observations of HD\,37061 strongly support this
hypothesis, as the measured circular polarisation seems roughly
proportional to the $Q/I$ value measured in the instrument reference
system (Fig.~\ref{Fig_XTalk}).  This phenomenon is discussed by
Bagnulo et al.\ (2009). It can be physically ascribed either to the
instrument collimator or the LADC. If cross-talk is stable, the
polarisation intrinsic to the source can be obtained by averaging the
signals measured at two instrument PAs that differ by 90\degr, i.e.,
using exactly the same method adopted for linear polarisation. For
HD\,37061, with grism 600\,B, the average signal is $ (P_V^{(30)} +
P_V^{(120)})/2 \sim 0.03$\,\% (Fig.~\ref{Fig_CIR_IP}).  By combining
the pairs of observations at 120\degr\ and 210\degr\ we find $\sim
0.1$\,\%. Therefore the cross-talk from linear to circular
polarisation is not constant with time, telescope or instrument
position. In grisms 1200\,R we obtain $(P_V^{(30)} + P_V^{(120)})/2
\sim 0.03$\,\%. In Fig.~\ref{Fig_CIR_IP} we also show the measurements
in grism 600\,B towards HD\,35149 giving $(P_V^{(0)} + P_V^{(90)})/2
\sim 0.05$\,\% and towards HD\,93250 of $\sim 0.02$\,\%, respectively.

Our observing strategy reduces the cross-talk from linear to circular
polarisation. Nevertheless, the instrumental issues
require a more accurate calibration. We are able to achieve in the
continuum of the circular polarisation spectrum an accuracy of $\sim
0.03$\,\%. This is, however, insufficient to test the theoretical
predictions computed in Fig.~\ref{Polab.fig}. Linear polarisation
spectra that are corrected for instrumental signatures of the stars
HD\,37061 (Fig.~\ref{HD37061.fig}), HD\,93250
(Fig.~\ref{HD93250.fig}), HD\,99872 (Fig.~\ref{HD99872.fig}), and
HD\,37903 (Fig.~\ref{exthd37903.fig}) are discussed below.


\section{Fitting results}

The dust model is applied to average properties of the ISM and towards
specific sight-lines. We set-up models so that abundance constraints
are respected to within their uncertainties (Sect.~\ref{abu.sec}), and
we fit extinction, polarisation and emission spectra.  The observed
extinction is a line-of-sight measurement to the star, while the IR
emission is integrated over a larger solid angle and along the
    entire line-of-sight through the Galaxy. In principle both
measurements treat different dust column densities.  Therefore an
extra assumption is made that the dust responsible for extinction and
emission has similar physical characteristics.  Dust emission from
dense and cold background regions may give a significant contribution
to the FIR/submm. However, PAH are mostly excited by UV photons that
cannot be emitted far away from the source, and the same holds for
warm dust that needs heating by a nearby source.

\subsection {Dust in the solar neighbourhood \label{ism.sec}}

\begin{figure}

{\vspace{-0.35cm} 
\hspace{-0.075cm}   \includegraphics[width=9.2cm]{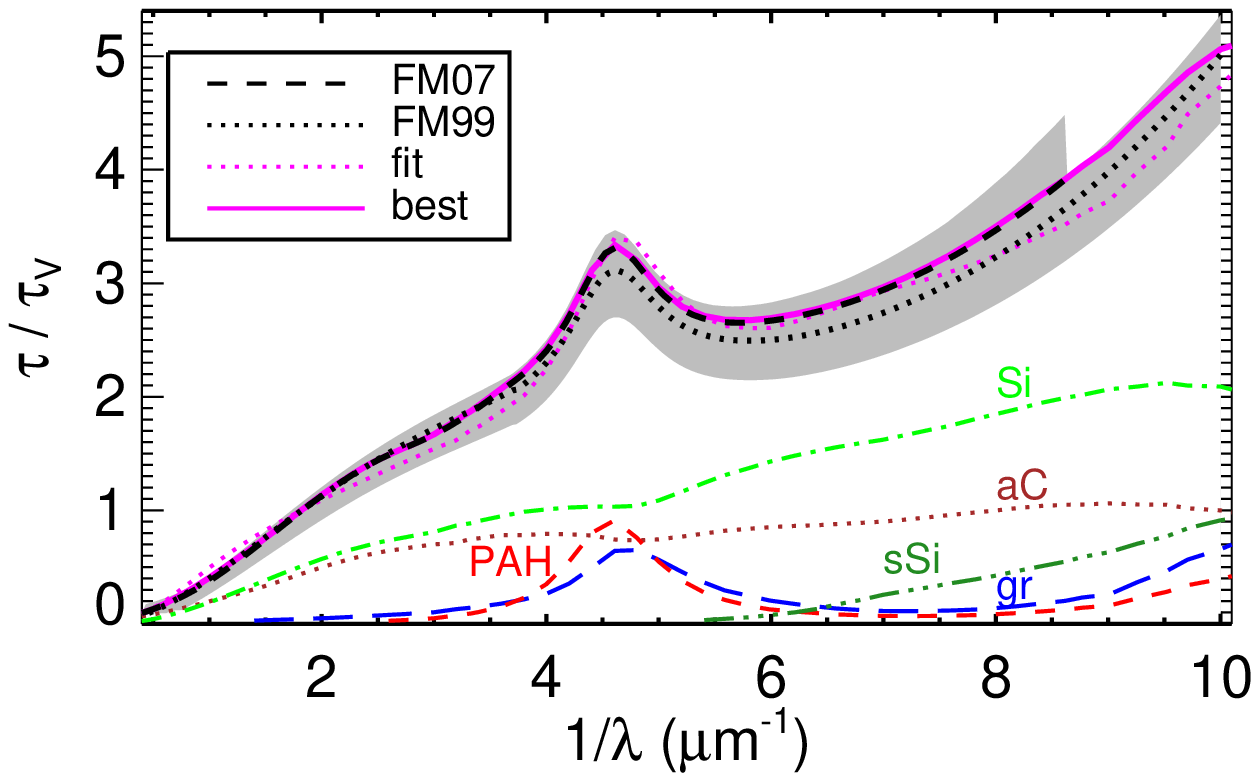}}

\vspace{-0.25cm} 
{\hspace{-0.075cm}   \includegraphics[width=9.2cm]{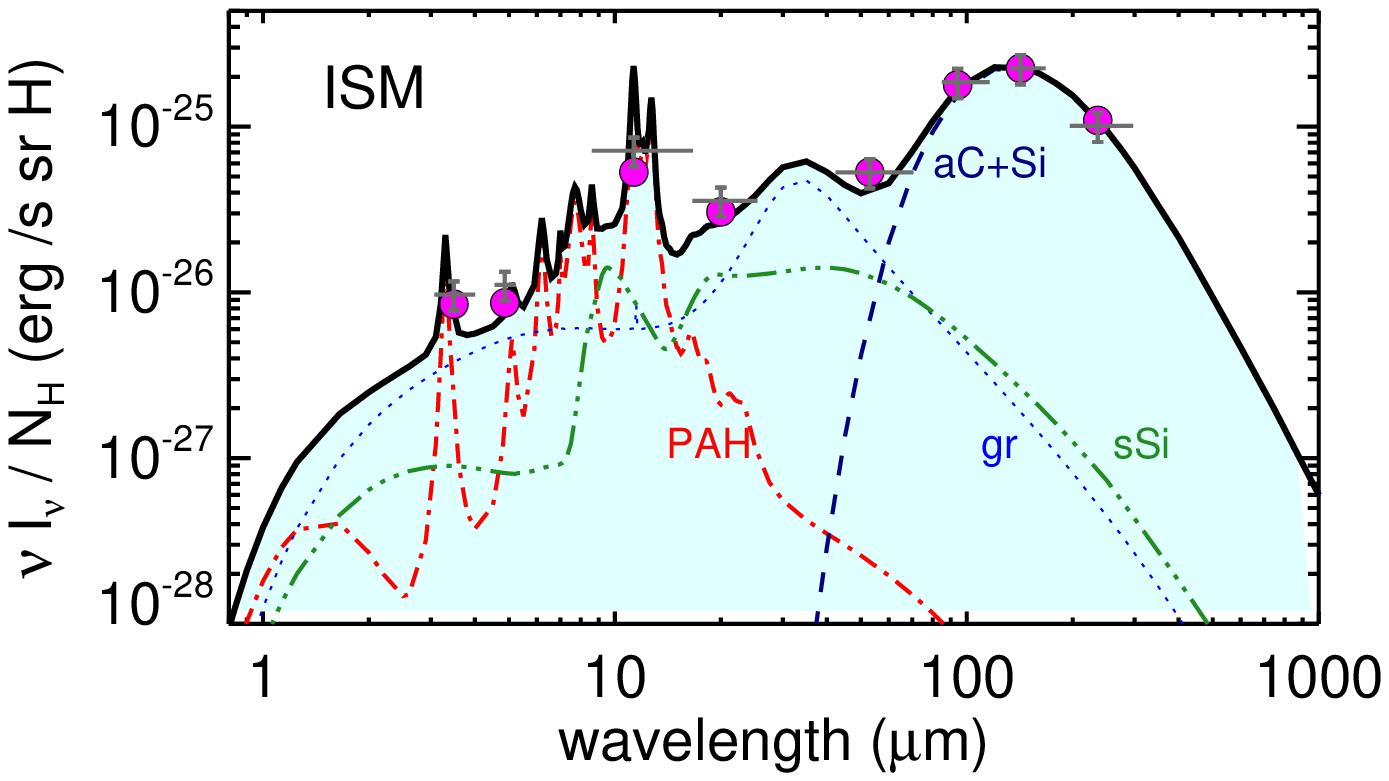}}

{\hspace{-0.075cm}   \includegraphics[width=9.2cm]{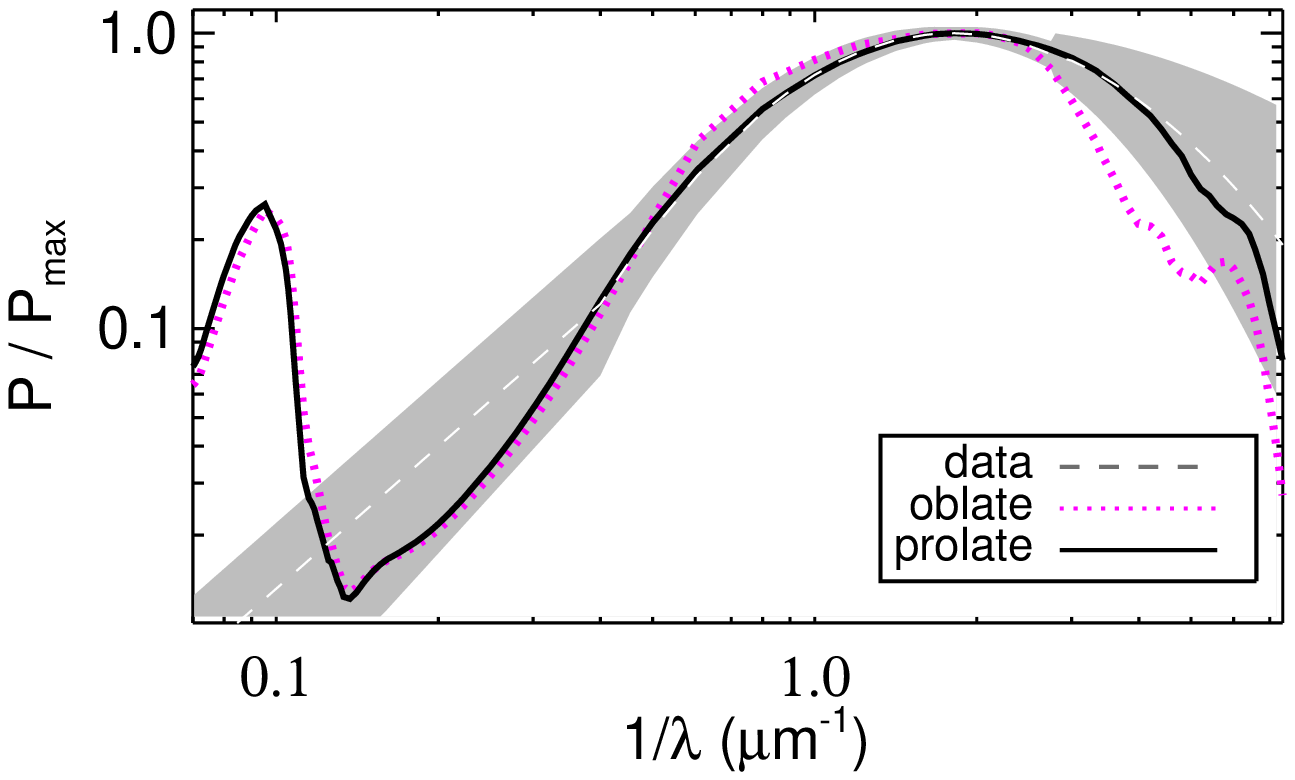}}
{\vspace{-0.25cm} 
  \caption{Dust in the solar neighbourhood. Mean (dashed) and
    1$\sigma$ variation (hatched area) of the observed extinction
    curves in the ISM, (up to 8.6\,$\mu$m$^{-1}$ by Fitzpatrick (1999)
    and $\le 10$\,$\mu$m$^{-1}$ by Fitzpatrick \& Massa (2007)).  The
    contribution of the individual dust components to the total
    extinction of the model with scaled (magenta line) and unscaled
    (magenta dotted) cross sections (Eq.~\ref{f.eq}) are given (top).
    Emission normalized per H atom when dust is heated by the
    ISRF. High Galactic latitudes observations with $1\sigma$ error
    bars (gray) from DIRBE (Arendt et al. 1998) and FIRAS (Finkbeiner
    et al. 1999). The model fluxes convolved with the band passes of
    the observations are shown as filled circles.  The contribution of
    the dust components to the total emission (black line) is shown
    (middle).  Mean (dashed) and 1$\sigma$ variation (hatched area) of
    the observed linear polarisation normalised to the maximum
    polarisation as given by Voshchinnikov et al. (2012). The
    normalised linear polarisation of silicates with prolate (black
    line) and oblate (magenta dotted) shapes is shown
    (bottom).  \label{ism.fig}}}
\end{figure}


The average of the extinction curves over many sight lines is taken to
be representative for the diffuse ISM of the Milky Way and gives
$R_{\rm{V}} = 3.1$.  Such average extinction curves and their scatter
are given by Fitzpatrick (1999) up to 8.6$\mu$m$^{-1}$ and Fitzpatrick
\& Massa (2007) up to 10$\mu$m$^{-1}$.  They are displayed in
Fig.~\ref{ism.fig} as a ratio of the optical depths. The average
extinction curve is approximated by varying the exponent of the dust
size distribution $q$ and the relative weights $w_i$ of the dust
populations. The upper size limit of large grains is derived by
fitting the mean polarisation spectrum of the Milky Way discussed
below. We find $r_{+} = 0.45\mu$m. In 1\,g of dust we choose: 546\,mg
to be in large silicates, 292\,mg in amorphous carbon, 43\,mg in
graphite, 82\,mg in small silicates, 14\,mg in small and 23\,mg in
large PAHs, respectively. A fit that is consistent to within the
errors of the mean extinction curve is derived with the parameters of
Table~\ref{para.tab} and is shown in Fig.~\ref{ism.fig}.

Comparing results of the extinction fitting with infrared observations
at high galactic latitudes has become a kind of benchmark for models
that aim to reproduce dust emission spectra of the diffuse ISM in the
Milky Way (D\'{e}sert et al. 1990; Siebenmorgen \& Kr\"ugel 1992; Dwek
et al. 1997; Li\& Draine 2001; Compiegne et al. 2011; Robitaille et
al. 2012).  In these models the dust is heated by the mean intensity,
$J_{\nu}^{\rm {ISRF}}$, of the interstellar radiation field in the
solar neighbourhood (Mathis et al. 1983). Observations at Galactic
latitude $\vert b \vert \ga 25\degr$ using DIRBE (Arendt et al. 1998)
and FIRAS (Finkbeiner et al. 1999) on board of COBE are given in
$\lambda I_{\lambda} /N_{\rm H}$ (erg/s/sr/H-atom), with hydrogen
column density $N_{\rm H}$ (H-atom/cm$^2$). Therefore we need to
convert the units and scale the dust emission spectrum computed by
Eq.~(\ref{emis.eq}) by the dust mass $m_d$ (g--dust/H-atom); this
gives $I_{\lambda} /N_{\rm H} = m_d \ \epsilon_{\lambda}$.


\begin{figure}
{\hspace{-0.75cm}   \includegraphics[width=10.cm]{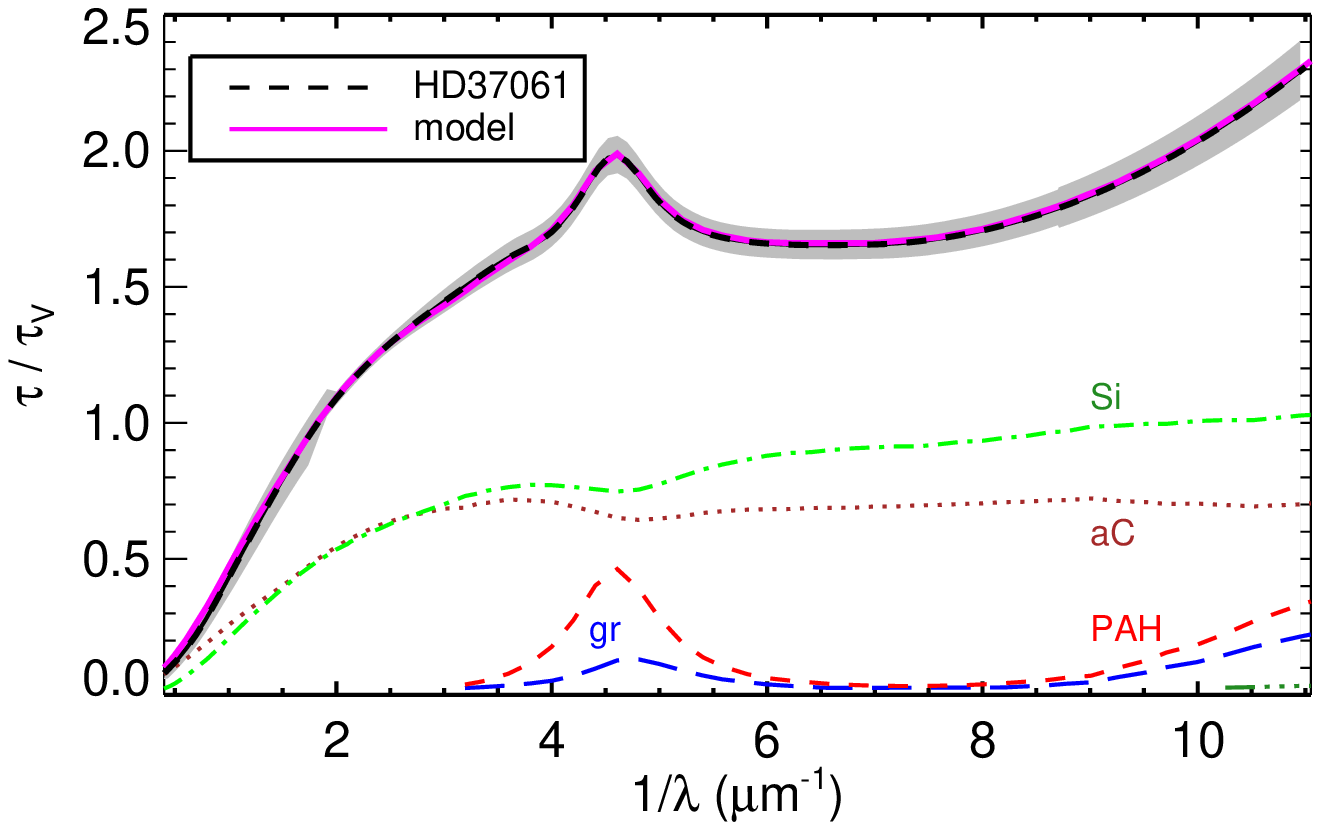}}

{\hspace{-0.75cm}    \includegraphics[width=10.cm]{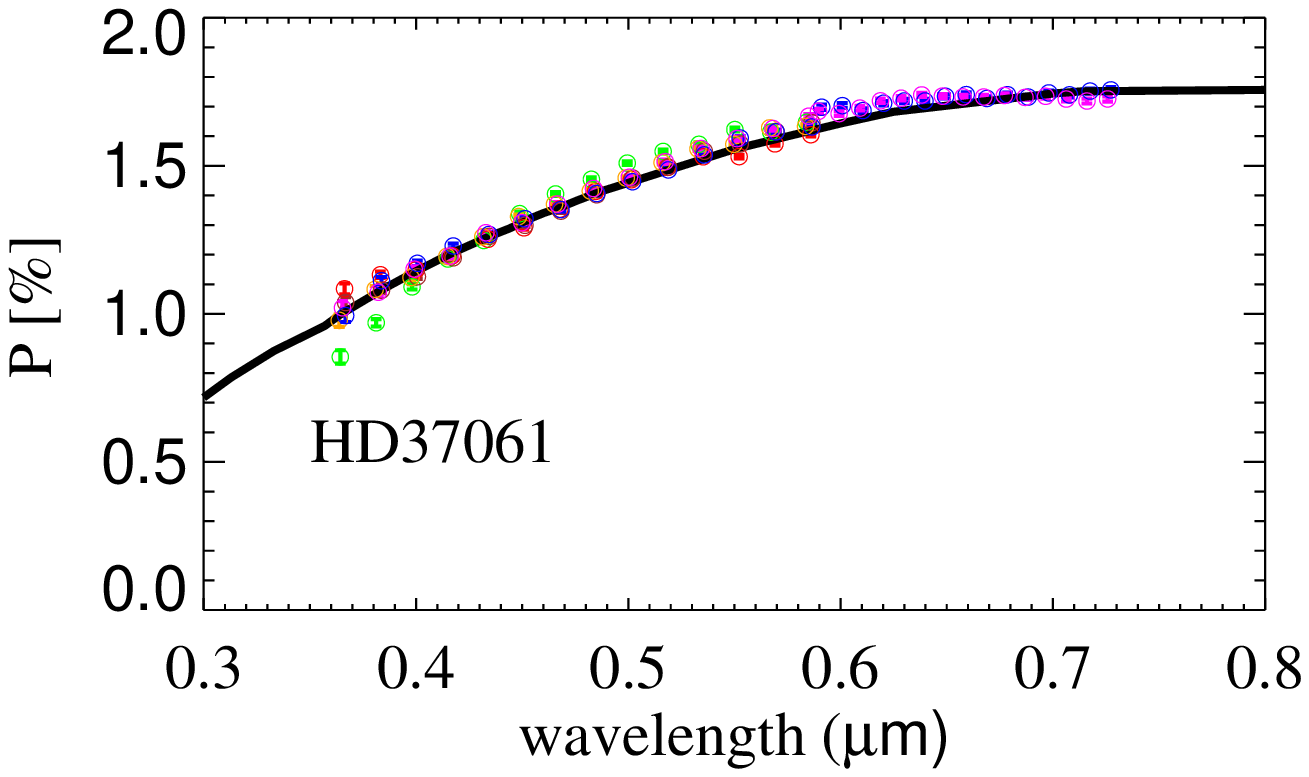}}

\caption{\label{HD37061.fig} Extinction curve (top) and polarised
  spectrum (bottom) of HD\,37061.  The observed extinction curve
  (black dashed line) with 1\,$\sigma$ error bars (hatched area) and
  the dust model (magenta solid line) with contribution from
  individual dust components are shown as labelled. The FORS2 linear
  polarisation spectra (circles) and the model (black solid line) is
  shown. Model parameters are given in Table~\ref{para.tab}.  }
\end{figure}

\begin{figure}  [htb]
{\hspace{-0.75cm}    \includegraphics[width=10.cm]{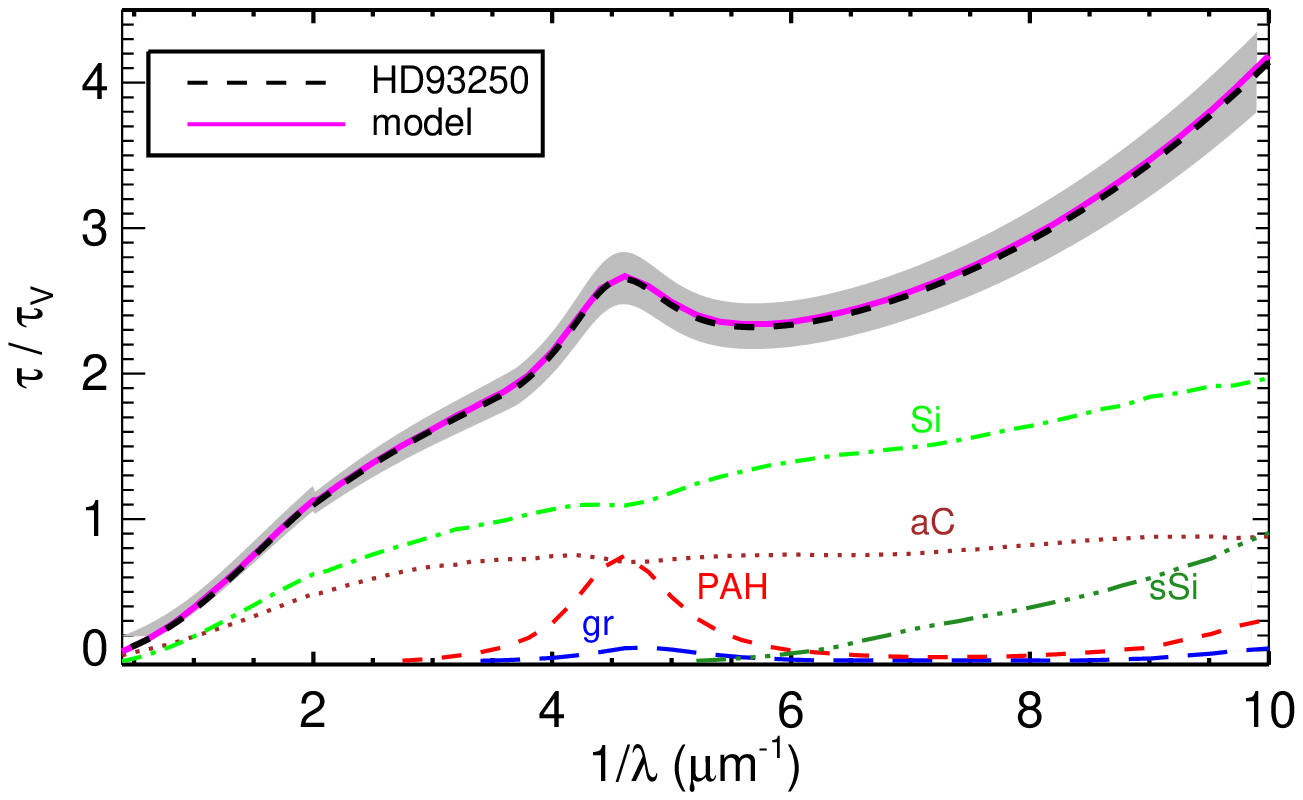}}

{\hspace{-0.75cm}    \includegraphics[width=10.cm]{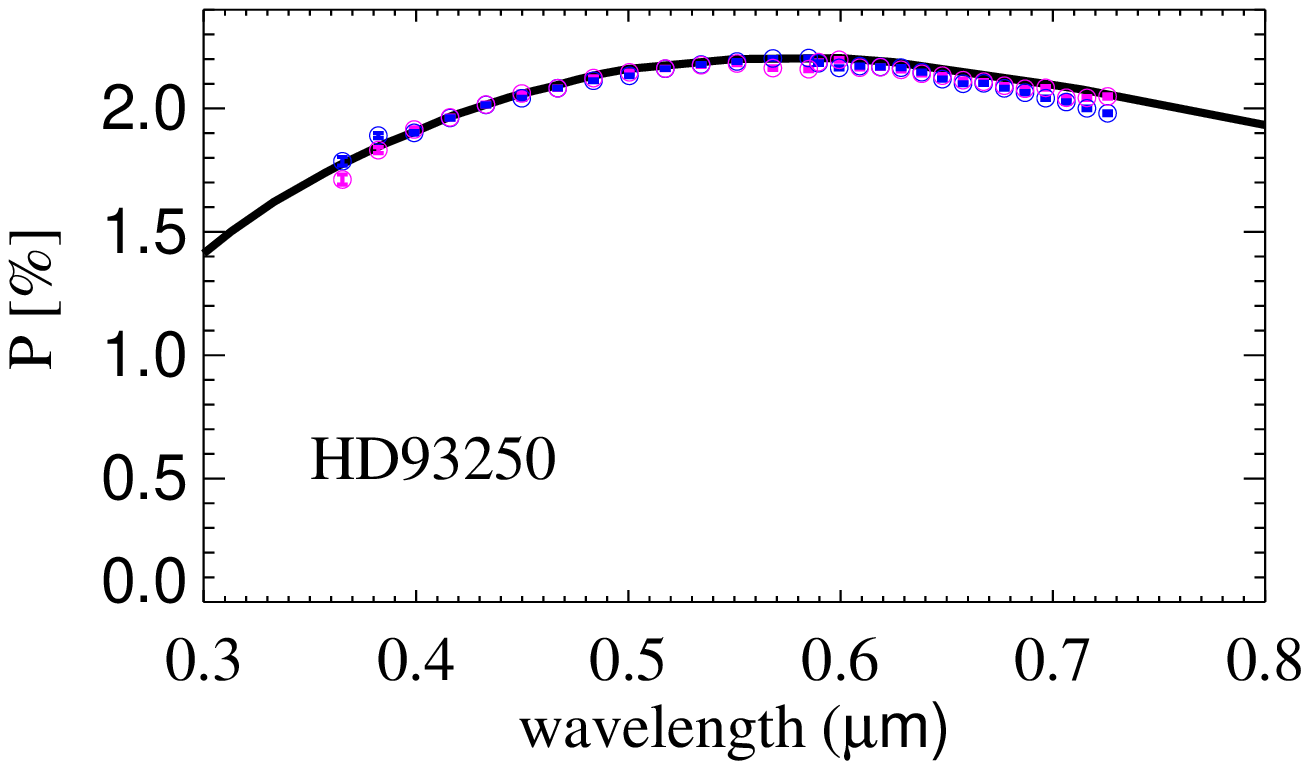}}

   \caption{\label{HD93250.fig} Same as Fig.\ref{HD37061.fig} for HD\,93250.}
\end{figure}

We derive the conversion factor by matching the model flux to that in
the 140$\mu$m DIRBE band pass. This gives $m_{\rm d}= 1.48 \times
10^{-26}$ (g-dust/H-atom) and a gas--to--dust mass ratio towards that
direction of $(1.36 \ m_{\rm p}/m_{\rm d}) \sim 153$. If one corrects
for the different specific densities $\rho_i$ of the dust materials,
our estimate is consistent within 8\% of that by Li \& Draine
(2001). A Kramers--Kronig analysis is applied by Purcell (1969)
finding as principle value an upper limit of $1.36 \ m_{\rm p}/m_{\rm
  d} <170$, where a specific density of the dust material of $\rho
\leq 2.5$\,(g/cm$^3$) and a correction factor of 0.95 for the grain
shape is assumed (cmp. Eq. (21.17) in Draine 2011).  The dust mass in
the model shall be taken as a lower limit because there could be
undetected dust components made up of heavy metals. For example, a
remaining part of Fe that is not embedded into the amorphous olivines
(Voshchinnikov et al. 2012) might be physically bonded in layers of
iron-fullerene clusters (Fe$\cdot$C$_{60}$, Lityaeva et al., 2006) or
other iron nanoparticles (Draine \& Hensley 2013). To our knowledge
there is no firm spectral signature of such putative components
established so we do not consider them here. Nevertheless, in the
discussion of the uncertainty of $m_d$ one should consider the
observational uncertainties in estimates of $N_{\rm H}$ towards that
region as well.

The total emission and the spectrum of each grain population is shown
in Fig.~\ref{ism.fig}. All observed in-band fluxes are fit within the
uncertainties.  We compare the dust emission computed by applying
cross sections of the initial physical model with that of the
fine-tuned cross sections (Eq.~\ref{f.eq}). We find that the
difference of these models in the DIRBE band pass is less than
1\%. The graphite emission peaks in the 20--40$\mu$m region. This
local maximum in the emission is due to the optical constants. We
verified that the emission by graphite in that region becomes flatter
when applying different optical constants of graphite, such as those
provided by Laor \& Draine (1993) and Draine \& Lee (1984). Such
flatter graphite emission is shown by e.g., Siebenmorgen \& Kr\"ugel
(1992). The 12$\mu$m DIRBE band is dominated by the emission of the
PAHs.  We have difficulties to fit this band by adopting PAH cross
sections as derived for starburst nuclei.  Li \& Draine (2001)
underestimate the emission in this DIRBE band by 40\%. In their model
the 11.3 and 12.7$\mu$m bands do not depend on the ionisation
degree. In the laboratory it is observed that the ratio of the C-H
stretching bands, at 11.3 and 12.7$\mu$m, over the C=C stretching
vibrations, at 6.2 and 7.7$\mu$m, decrease manifold upon ionisation
(Tielens\ 2008). Therefore in the diffuse ISM we vary the PAH cross
section as compared to the ones derived in the harsh environment of OB
stars where PAHs are likely to be ionised (Table~\ref{pah.tab}). Data
may be explained without small silicates.

\noindent
In the optical and near IR the observed polarisation spectra of the
ISM can be fit by a mathematical formulae, known as the Serkowski
(1973) curve:

\begin{equation}
\frac{p(\lambda)}{p_{\max}} = \exp \left[ -k_p \ \ln^2
    \left( \frac{\lambda_{\max}}{\lambda} \right) \right]\,,
\label{serk.eq}
\end{equation}

\begin{figure} [h!tb]
{\hspace{-0.75cm}    \includegraphics[width=10.cm]{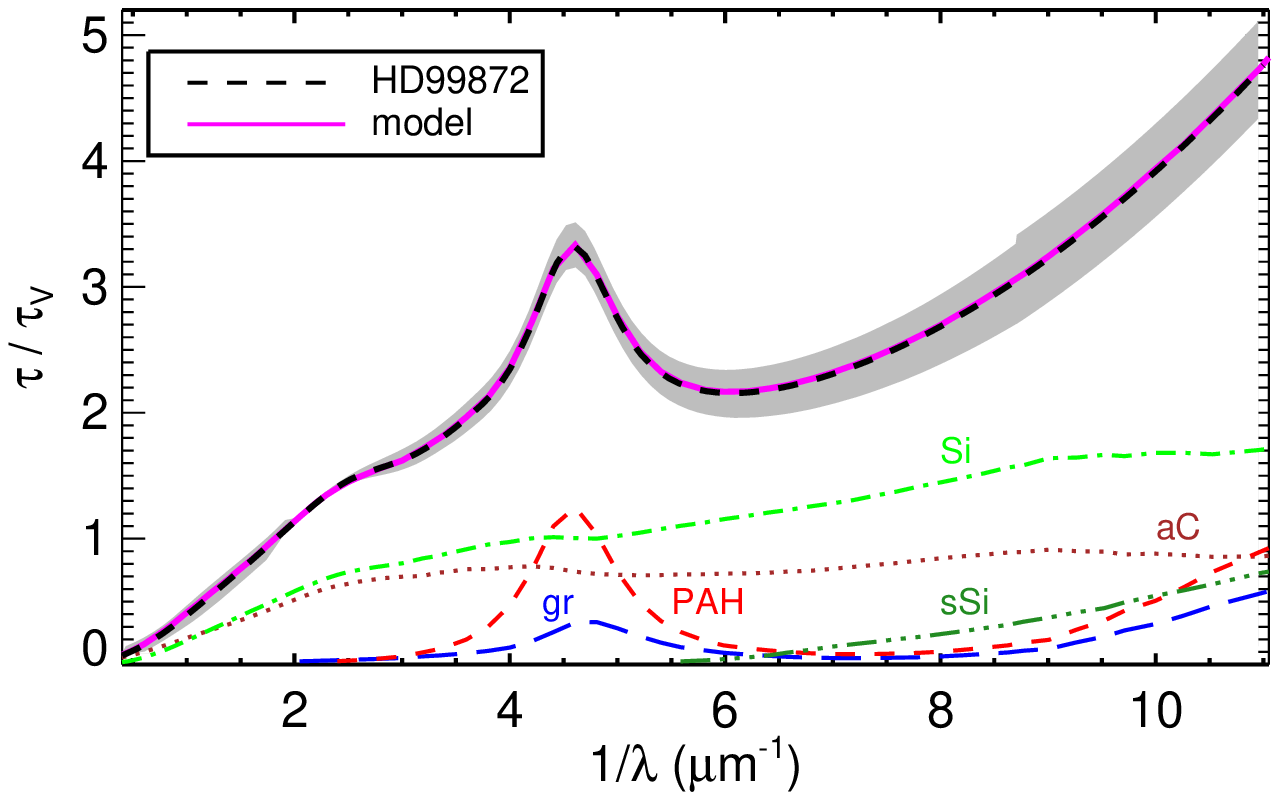}}

{\hspace{-0.75cm}    \includegraphics[width=10.cm]{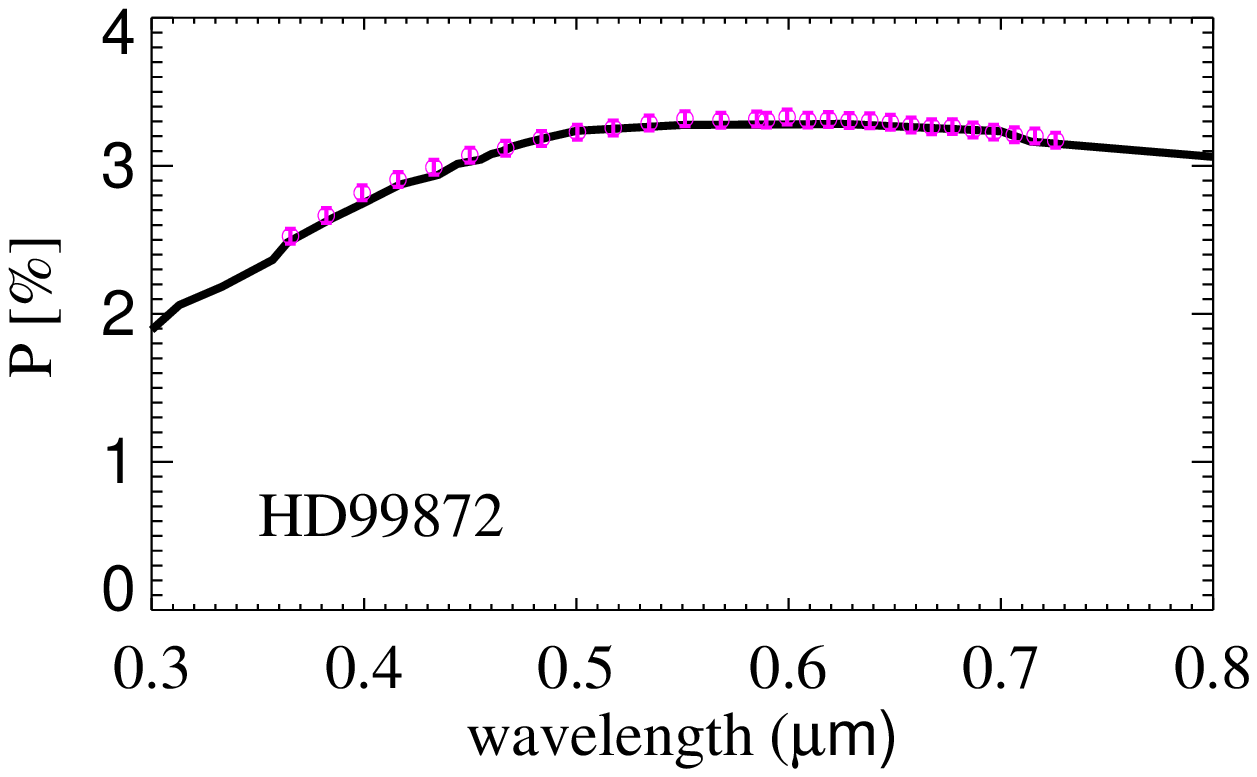}}

{\hspace{-0.75cm}    \includegraphics[width=10.cm]{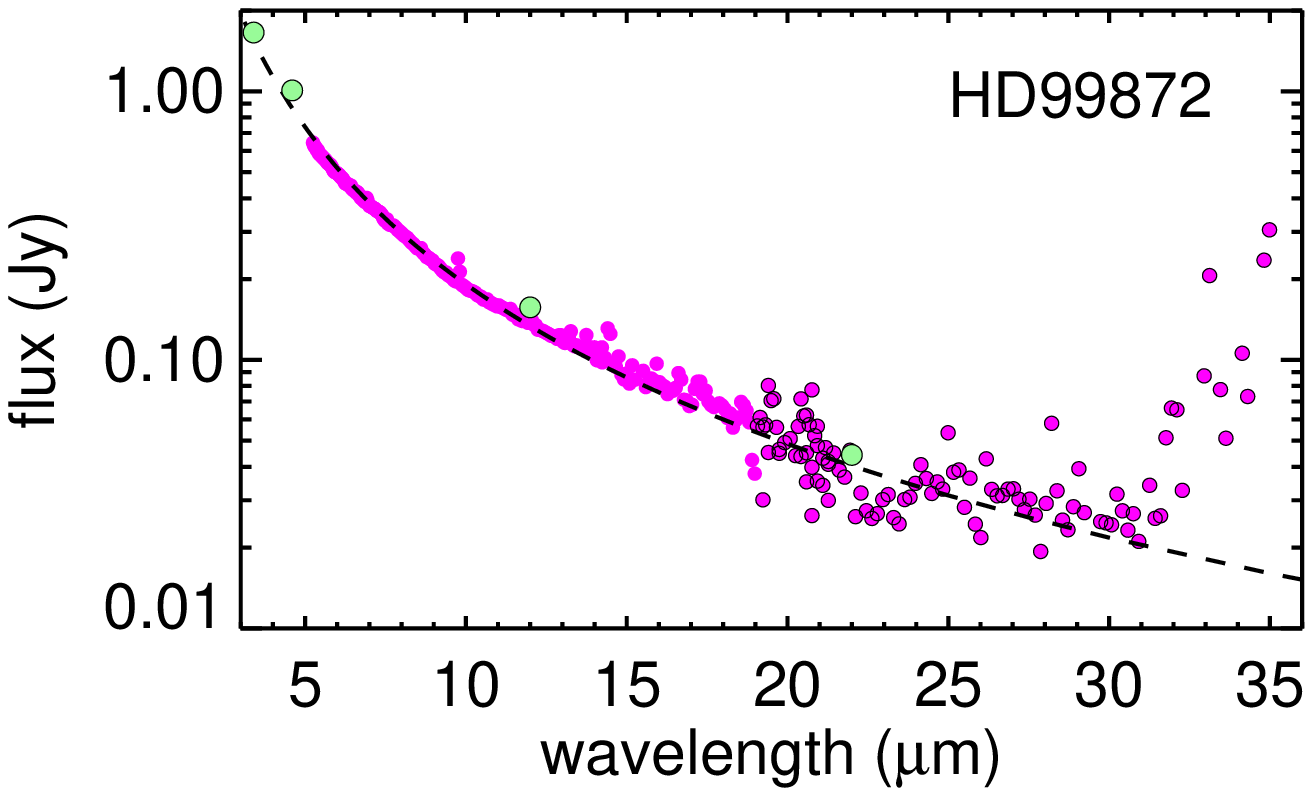} }

\caption{\label{HD99872.fig} Top and middle panels as in
  Fig.\ref{HD37061.fig} for HD\,37061.  Bottom: 3 -- 36\,$\mu$m
  emission observed with Spitzer/IRS (red filled circles) and WISE
  (green filled circles). The photospheric emission of the star is
  represented by the dashed line}
\end{figure}

\noindent
where the maximum polarisation is observed to be $p_{\rm {max}}/A_V
\la 3\%/$mag (Whittet 2003). In the thermal IR at $\lambda >
2.5\,\mu$m polarisation data are fit by a power--law where $p(\lambda)
\propto \lambda^{-t}$ with $1.6 \le t \le 2$ (Martin et al., 1992,
Nishiyama et al. 2006). This fit naturally breaks down in the 10$\mu$m
silicate band. The average observed linear polarisation of the ISM is
displayed in Fig.~\ref{ism.fig}.  The Serkowski curve is fit without
carbon particles, only silicates are aligned. We find a good fit
assuming that silicates with particle sizes of radii between 100 --
450\,nm are aligned. Below 0.3$\mu$m the mean polarisation spectrum is
better explained by dust of prolate than oblate structure
(Fig.~\ref{ism.fig}).

\subsection {HD\,37061}

We repeat the exercise of the Sect.~\ref{ism.sec} and model
extinction, polarisation and, when available, dust emission spectra.
So far we modelled dust properties from extinction and polarisation
data when observations are averaged over various sight lines. In the
following we fit data towards a particular star and choose those for
which we present observations of the linear polarisation spectrum.

This star is of spectral type B1.5V, and located at 720\,pc from us.  The
extinction curve is compiled by Fitzpatrick \& Massa (2007). The
selective extinction is $R_{\rm V} = 4.55 \pm 0.13$ and the visual
extinction $A_{\rm V} = 2.41 \pm 0.11$ (Voshchinnikov et al.
2012).  ISO and Spitzer spectra of the dust emission are not
available. Polarisation spectra are observed by us with
of FORS/VLT. The spectra are consistent with earlier measurements of
the maximum linear polarisation of $p_{\rm {max}} = 1.54 \pm 0.2$\,\%
at $\lambda_{\rm {max}} = 0.64 \pm 0.04\,\mu$m by Serkowski et
al. (1975). A fit to the extinction curve and the polarisation
spectrum is shown in Fig.~\ref{HD37061.fig}.  The observed
polarisation spectrum is fit by silicate grains that are of prolate
shape with IDG alignment and $\Omega \sim 55^{\rm{o}}$, other
parameters as of Table~\ref{para.tab}.

\subsection {HD\,93250}

This star is of spectral type O6V and located 1.25\,kpc from us.  The
extinction curve is compiled by Fitzpatrick \& Massa (2007) and
between $3.3\mu \rm{m}^{-1} \la \lambda^{-1} \la 11\mu \rm{m}^{-1}$ by
Gordon et al. (2009), who present spectra of the Far Ultraviolet
Spectroscopic Explorer (FUSE) and supplemented spectra from the
International Ultraviolet Explorer (IUE). The selective extinction is
$R_{\rm V} = 3.55 \pm 0.34$ and the visual extinction $A_{\rm V} =
1.54 \pm 0.1$ (Gordon et al.\ 2009).  Polarisation spectra are
observed by us in two orientations of the instrument; other
polarisation data as well as ISO or Spitzer spectra are not
available. A fit to the extinction curve and the polarisation spectrum
is shown in Fig.~\ref{HD93250.fig}. Dust parameters are summarised in
Table~\ref{para.tab}.


\subsection {HD\,99872}
The star has spectral type B3V and is located at 230\,pc from us.  The
extinction curve is compiled by Fitzpatrick \& Massa (2007) and
between $3.3\mu \rm{m}^{-1} \la \lambda^{-1} \la 11\mu \rm{m}^{-1}$ by
Gordon et al. (2009). The selective extinction is $R_{\rm V} = 2.95
\pm 0.44$ and the visual extinction $A_{\rm V} = 1.07 \pm 0.04$
(Gordon et al.\ 2009).  The spectral shape of the FORS polarisation
can be fit by adopting aligned silicate particles with a prolate
shape, while a contribution of aligned carbon grains is not required.
The observed maximum polarisation of 3.3\,\% is reproduced assuming
IDG alignment with efficiency computed by Eq.(\ref{align}) and $a/b
\sim 6$. Dust parameters are summarised in Table~\ref{para.tab}. A fit
to the extinction curve, polarisation spectrum, and IR emission is
shown in Fig.~\ref{HD99872.fig}.  The IR emission is constrained by
WISE (Cutri et al.\ 2012) and by photometry and a Spitzer/IRS archival
spectrum (Houck et al.\ 2004). The 3 -- 32$\mu$m observations are
consistent with a 20000\,K blackbody stellar spectrum, and do not
reveal spectral features by dust.  At wavelengths $\ga 32\,\mu$m the
observed flux is in excess of the photospheric emission. The excess
has such a steep rise that it is not likely to be explained by
free--free emission. It is pointing towards emission from a cold dust
component such as a background source or a faint circumstellar dust
halo. The optical depth of such a putative halo is too small to
contribute to the observed polarisation that is therefore of
interstellar origin (Scicluna et al.\ 2013). Unfortunately it has not
been possible to find in the literature high quality data at longer
wavelengths to confirm and better constrain the far IR excess.


\section {The Reflection Nebula NGC\,2023 \label{hd37903}}

We present observations towards the star HD\,37903, which is the primary
heating source of the reflection nebula NGC\,2023. It is located in the
Orion Nebula at a distance of $\sim 400$\,pc (Menten et al. 2007). The
star is of spectral type B1.5V. It has carved a quasi--spherical
dust--free HII region of radius of $\leq 0.04$\,pc (Knapp et al.\
1975, Harvey et al.\ 1980). Dust emission is detected further out, up
to several arcmin, and is distributed in a kind of bubble--like
geometry (Peeters et al. 2012). In that envelope ensembles of dust
clumps, filaments and a bright southern ridge are noted in near IR
and HST images (Sheffer et al.\ 2011).

\begin{figure} [h!tb]
\hspace{-0.cm}
{\hspace{-0.75cm}     \includegraphics[width=10.cm]{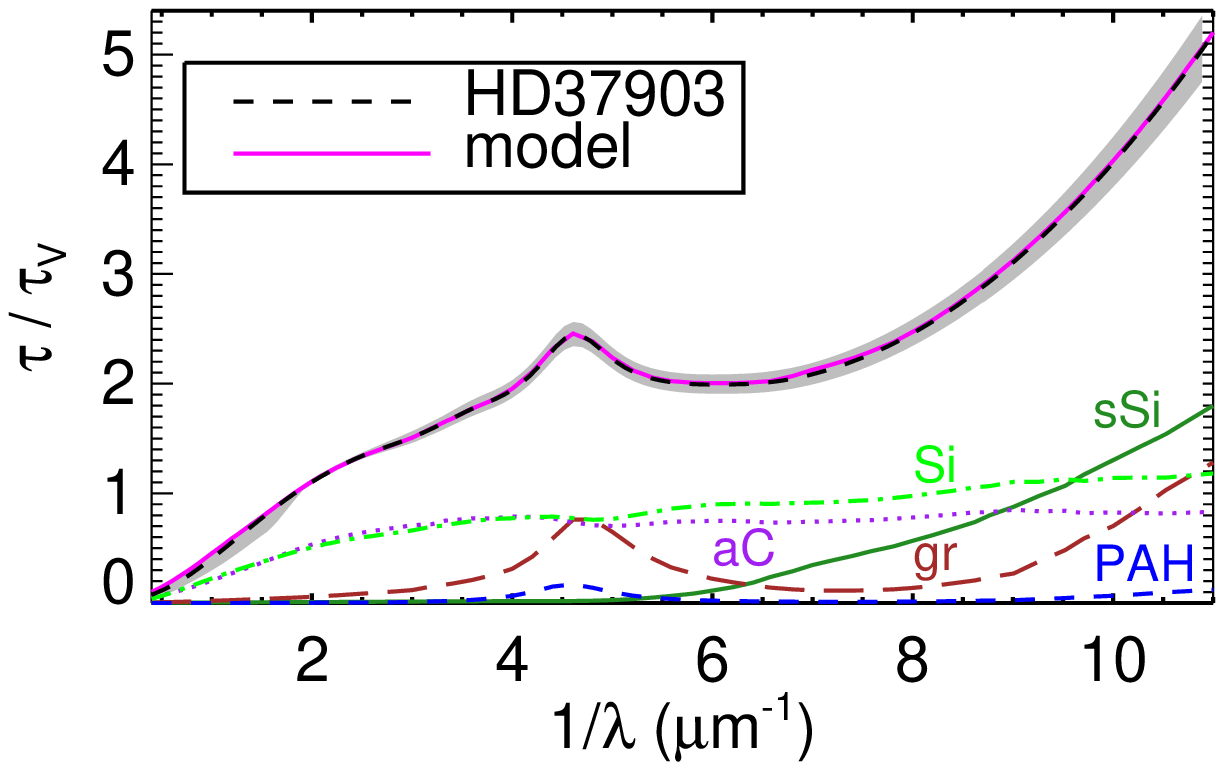}}

{\hspace{-0.75cm}     \includegraphics[width=10.cm]{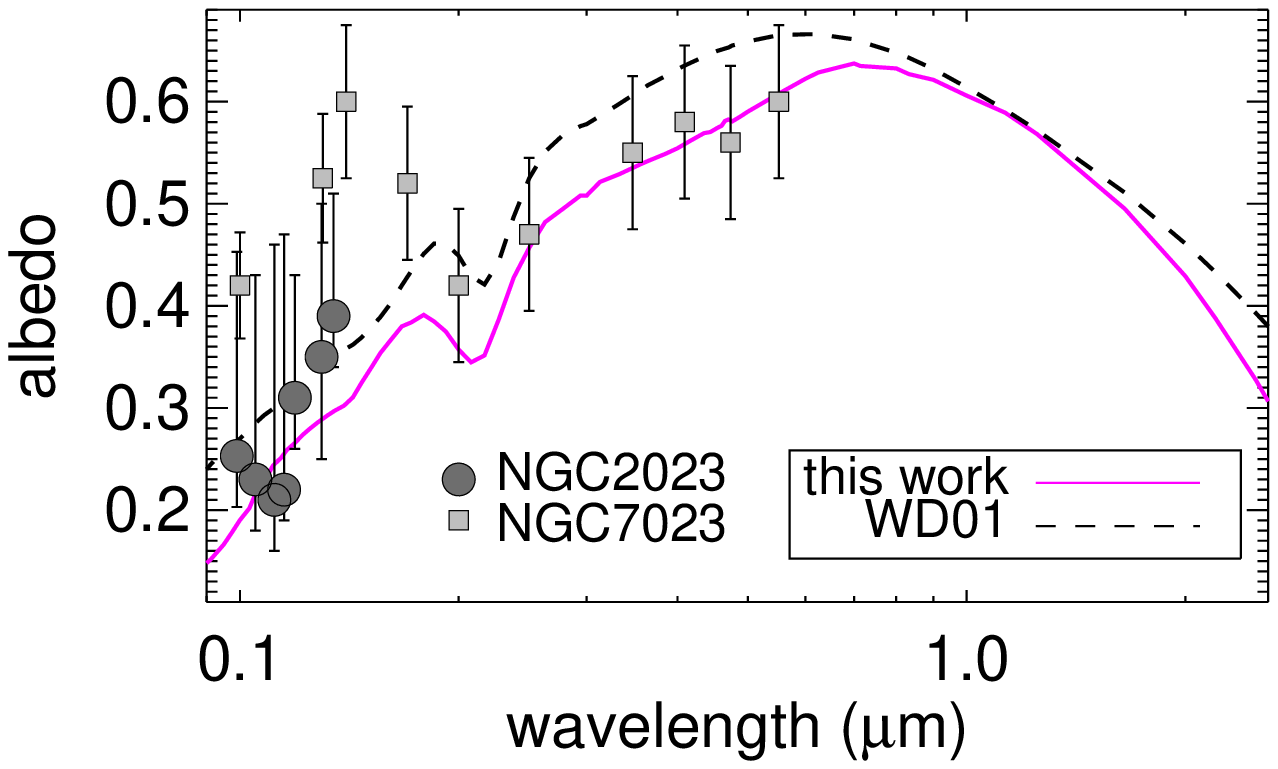}}

{\hspace{-0.75cm}     \includegraphics[width=10.cm]{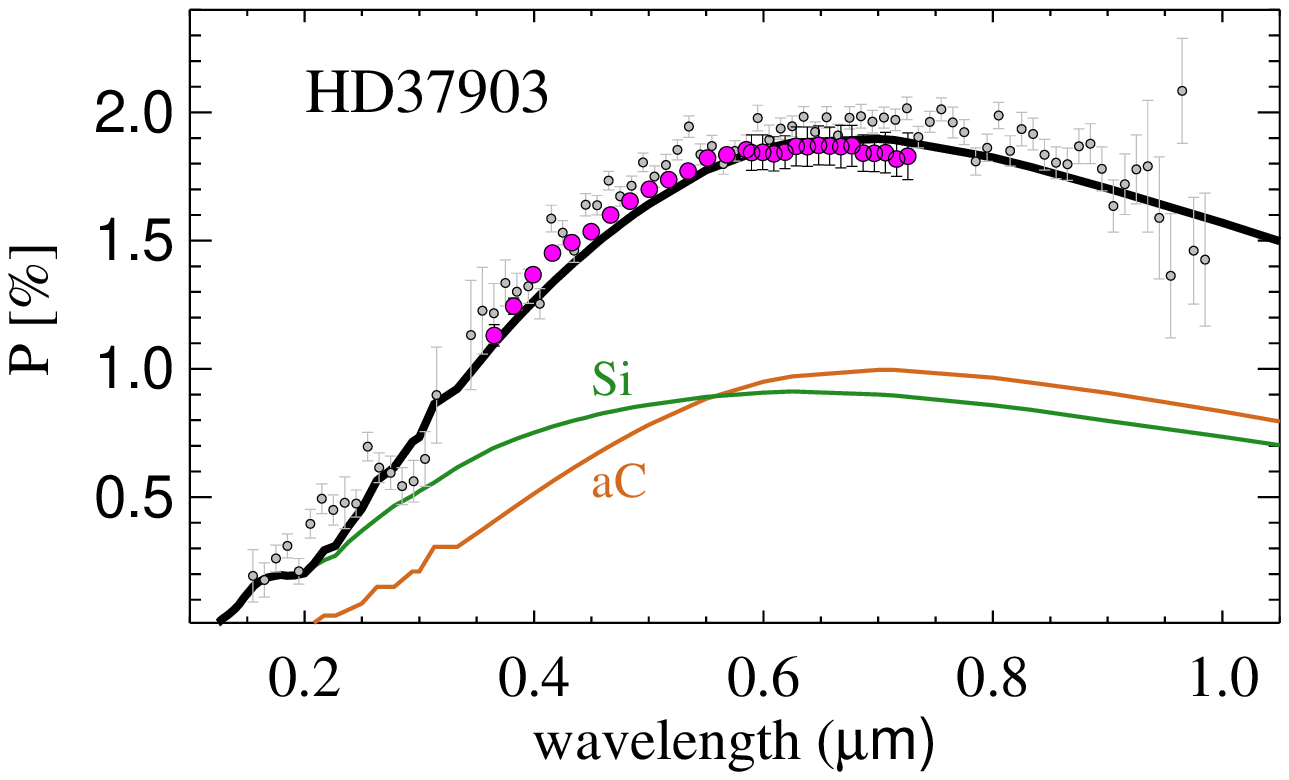}}

\caption{Extinction (top, with notation of Fig.~\ref{HD37061.fig}.),
  albedo (middle) and linear polarisation spectrum (bottom) of
  HD\,37903.  The albedo of the dust model of this work (magenta line)
  and by Weingartner \& Draine (2001, dashed) is displayed together
  with estimates of the dust albedo of the reflection nebulae NGC\,7023
  (square) and NGC\,2023 (circle).  The observed polarisation spectrum
  with 1$\sigma$ error bars of this work (red symbols) and as compiled
  by Efimov (2009, gray symbols) is displayed together with the model
  (black line) that is the sum of aligned silicates (green) and
  amorphous carbon (orange) particles. Dust parameters are summarised
  in Table~\ref{para.tab}. \label{exthd37903.fig}}
\end{figure}


\begin{figure}  [h!tb]

{\hspace{-0.7cm}   \includegraphics[width=10cm]{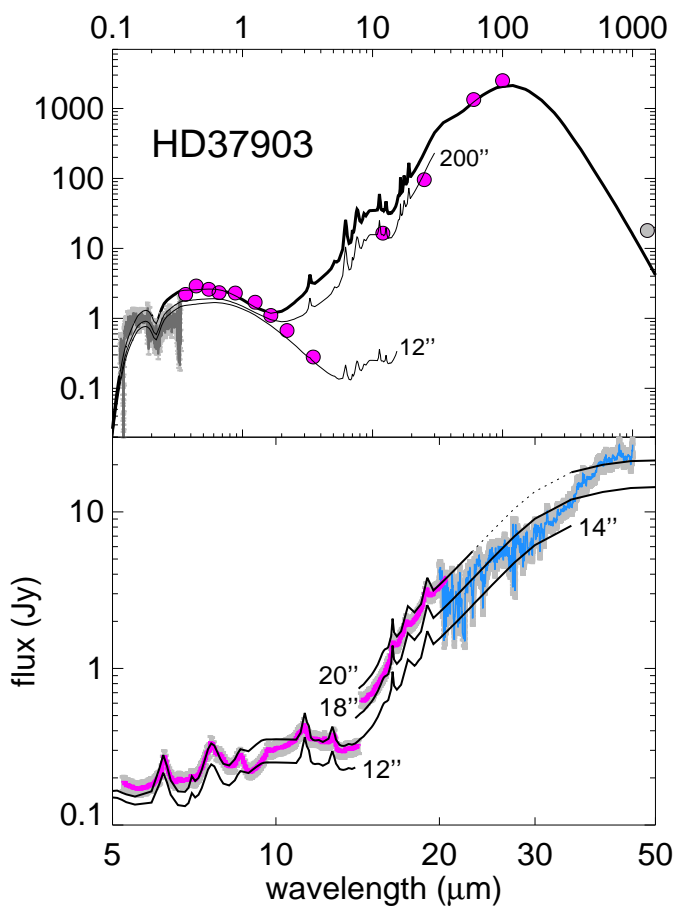}}

\caption{Spectral energy distribution of HD\,37903 that is heating
  NGC\,2023 (top) is shown together with a zoom into the 5 -- 50 $\mu$m
  region (bottom): photometric data (symbols); spectra by IUE (dark
  gray), Spitzer/IRS (magenta) and ISOSWS (blue) with 1$\sigma$ error
  (gray hatched area) and the model (black lines) in apertures as
  indicated.  \label{sedhd37903.fig} }
\end{figure}


\subsection {Extinction \label{hd37903dust.sec}}
The extinction curve towards HD\,37903 is observed by Fitzpatrick \&
Massa (2007) and Gordon et al. (2009). The extinction towards HD\,37903
is low and estimates range between A$_{\rm V} \sim 1.2 - 1.5$\,mag
(Burgh et al.\ 2002, Compiegne et al.\ 2008, Gordon et al.\ 2009). The
selective extinction is $R_{\rm V} = 4.11 \pm 0.44$ (Gordon et
al. 2009). We fit the extinction curve with parameters as of
Table.~\ref{para.tab}. The resulting fit together with the
contribution of the various grain populations is shown in
Fig.~\ref{exthd37903.fig}.

The total amount of energy removed by dust from the impinging light
beam is equal to (1-$\Lambda$), i.e. it is related to the particle
albedo $\Lambda$ (Eq.~\ref{albedo.eq}) and the angular distribution of
scattered light. The latter is determined by the asymmetry parameter
$g$.  The albedo and the asymmetry parameter cannot be derived
separately and are only observed in a combined form (Voshchinnikov,
2002). These two quantities have been estimated for several reflection
nebulae in the Galaxy (Gordon 2004).  Reflection nebulae are bright
and often heated by a single star. By assuming a simple scattering
geometry estimates of $\Lambda$ are given for the reflection nebula
NGC\,2023 (Burgh et al. 2002) and NGC\,7023 (Witt et al. 1982, Witt et
al. 1993). They are shown in Fig.~\ref{exthd37903.fig} together with
our dust model of HD\,37903 and for comparison the $R_{\rm V}=3.1$ model
by Weingartner \& Draine (2001). The albedo of both dust models are
similar. They are consistent within $1\sigma$ uncertainties of the
data of NGC\,2023, however deviate with the peak observed at 0.14$\mu$m
for NGC\,7023.

\subsection {Dust envelope}

PAH emission of NGC\,2023 is detected by Sellgren (1985), with ISOCAM by
Cesarsky et al. (2000) and Spitzer/IRS by Joblin et
al. (2005). Emission features at 7.04, 17.4 and 18.9\,$\mu$m are
detected and assigned to neutral fullerene (C$_{60}$) with an
abundance estimate of $<1$\% of interstellar carbon (Sellgren et
al. 2010). The fullerene show a spatial distribution distinct from
that of the PAH emission (Peeters et al. 2012). The dust emission
between 5 -- 35\,$\mu$m in the northern part of the nebula is modelled
by Compiegne et al. (2008), who apply a dust model fitting the mean
extinction curve of the ISM. The authors approximate that part of the
object in a plane parallel geometry and simplify the treatment of
scattering. Compiegne et al. (2008) find that the relative weight and
hence abundance of PAHs is five times smaller in the denser part of
the cloud than in the diffuse ISM.

We compile the spectral energy distribution (SED) towards HD\,37903
using photometry in the optical, UBVRI bands by Comeron (2003), in the
near-infrared, IJHKL filters by Burgh et al. (2002), IRAS (Neugebauer
et al. 1984; Joint IRAS Science 1994) and at 1.3mm by Chini et
al. (1984).  The SED is complimented by spectroscopy between
0.115$\,\mu$m and 0.32\,$\mu$m using data of the International Ultraviolet
Explorer (IUE), as available in the Mikulski archive for space
telescopes$^1${\footnote {http://archive.stsci.edu/iue/}}, the
Spitzer/IRS archival spectrum (Houck et al. 2004) and ISO/SWS
spectroscopy (Sloan et al. 2003). The observed SED is shown in
Fig.~\ref{sedhd37903.fig}.

The observed bolometric IR luminosity is used as an approximation of
the stellar luminosity of $L_* \sim 10^4$\Lsun . We take a blackbody
as stellar spectrum. Its temperature of $T_* = 21000$\,K is derived by
fitting the IUE spectrum and optical, NIR photometry first in a
dust-free model and correcting for a foreground extinction of $\tau
\sim D/1kpc = 0.4$.  Both parameters, $L_*$ and $T_*$, are appropriate
to the spectral type of HD\,37903. A weak decline of the dust density
with radius is assumed in the models by Witt et al. (1984).  We assume
in the nebulae a constant dust density of $\rho = 10^{-23}$\,g/cm$^3$
for simplicity.  The inner radius of the dust shell is set at
$r_{\rm{in}} = 10^{17}$\,cm. For the adopted outer radius of
$r_{\rm{out}} = 1$\,pc the total dust mass in the envelope is
$M_{\rm{dust}} = 0.62$\,\Msun \/ and the optical depth, measured
between $r_{\rm {in}}$ and $r_{\rm {out}}$, is $\tau_{\rm V}=0.8$. We
build up SEDs of the cloud within apertures of different angular
sizes. The models are compared to the observations in
Fig.~\ref{sedhd37903.fig}. The 1.3\,mm flux is taken as upper limit as
it might be dominated by cold dust emission of the molecular cloud
behind the nebula.  Model fluxes computed for different diaphragms
envelope reasonably well data obtained at various apertures.

Fullerenes are not treated as an individual dust component. The cross
sections of the PAH emission bands are varied to better fit the
Spitzer/IRS spectrum. They are listed in Table~\ref{pah.tab} and are
close to the ones estimated for starburst galaxies.  The cross
sections of the C-H bands at 11.3 and 12.7\,$\mu$m are much smaller
and of the C=C vibrations at 6.2 and 7.7\,$\mu$m slightly larger than
those derived for the solar neighbourhood. This finding is consistent
with a picture of dominant emission by ionised PAHs near B stars and
neutral PAHs in the diffuse medium. We note that alternatively, the
hydrogenation coverage of PAHs can be used in explaining a good part
of observed variations of PAH band ratios in various galactic and
extra--galactic objects (Siebenmorgen \& Heymann, 2012). We do not
find a good fit to the 10\,$\mu$m continuum emission if we ignore the
contribution of small silicate grains.


\begin{table*}[h!tb]
\begin{center}
  \caption {Parameters of the dust models. \label{para.tab}}
 \begin{tabular}{|l|c|c|c|c|c|}
\hline\hline
& & & & & \\
Parameter $^a$      & Solar & HD\,37061 & HD\,93250 & HD\,99872 & HD\,37903 \\
                    &neighbourhood &         &         &         &(NGC\,2023) \\
\hline
$r_+$ (nm)          & 440    &  485 &  380    &  400    & 485\\
$r_{\rm {-}}$ (nm)& 100    &  125 &  100    &  140    & 120\\
$q$                 & 3.4    &  3.0 &  3.3    &  3.3    & 3.2\\
$a/b$               &  2     &  2   &  2.2    &  6      &  2       \\
$w_{\rm {aC}}$      & 29.2   & 40.1 &  28.8   &  31.5   & 34.3\\
$w_{\rm {Si}}$      & 54.6   & 56.2 &  58.2   &  55.1   & 48.1\\
$w_{\rm {gr}}$      & 4.3    & 1.0    &  0.8    &  2.4    & 5.7\\
$w_{\rm {sSi}}$     & 8.2    & 0.3  &  8.9    &  5.4    & 11.2\\
$w_{\rm {PAHs}}$    & 1.4    & 0.8  &  1.3    &  2.4    & 0.2\\
$w_{\rm {PAHb}}$    & 2.3    & 1.6  &  1.9    &  3.2    & 0.5\\
\hline
\end{tabular}
\end{center}
{\bf {Notes.}} $^a$ Upper ($r_+$) and lower ($r_{\rm
  {-}}$) particle radius of aligned grains, exponent of the dust size
distribution ($q$), axial ratio of the particles ($a/b$), and relative
weight per g--dust (\%) of amorphous carbon ($w_{\rm{aC}}$), large
silicates ($w_{\rm{Si}}$), graphite ($w_{\rm{gr}}$), small silicates
($w_{\rm{sSi}}$), small ($w_{\rm{PAHs}}$) and large ($w_{\rm{PAHb}}$) PAHs,
respectively.

\end{table*}


\subsection {Polarisation}
Spectro--polarimetric data acquired with the polarimeters on board the
WUPPE and HPOL satellites{\footnote{www.sal.wisc.edu/}} and from
ground (Anderson et al. 1996) are compiled by Efimov (2009).  They
agree to within the errors of the polarisation spectra observed by us
with FORS2/VLT. We find a good fit to the polarisation spectrum only
when both silicates and amorphous carbon are aligned. The observed
polarisation spectrum is fit by grains that are of prolate shape,
$\Omega \sim 65\degr$ and other parameters as of Table~\ref{para.tab}.

As most of the extinction towards HD\,37903 is coming from the nebula,
one may also consider other alignment mechanisms of the grains such as
the radiative torque alignment proposed by Lazarian (2007).  Another
way to explain the observed polarisation could be due to dust
scattering. This would require significant inhomogeneities of the dust
density distribution of the nebula within the aperture of the
instruments. The contribution of scattering to the polarisation might
be estimated from a polarisation spectrum of a nearby star outside the
reflection nebula that is not surrounded by circumstellar dust.


\section{Conclusion \label{conclusion.sec}}
The main results of this paper are as follows: \\

{\bf 1)} We have presented an interstellar dust model that includes a
population of carbon and silicate grains with a power--law size
distribution ranging from the molecular domain (5\AA) up to 500\,nm.
Small grains are graphite, silicates and PAHs, and large spheroidal
grains are made of amorphous carbon and silicates.  The relative
weight of each dust component is specified, so that absolute
abundances of the chemical elements are not direct input parameters
and used as a consistency check (Eq.~\ref{w.eq}). We apply the
imperfect Davis--Greenstein alignment mechanism to spheroidal dust
particles, which spin and wobble along the magnetic field.  Their far
IR/submm absorption cross section is a factor 1.5 -- 3 larger than
spherical grains of identical volume. Mass estimates derived
    from submillimeter observations that ignore this effect are
    overestimated by the same amount.  The physical model fits
observed extinction curves to within a few percent and a perfect match
is found after fine adjustment of the computed cross sections
(Eq.~\ref{f.eq}).

{\bf 2)} The wavelength-dependent absorption cross-section of PAHs have
  been revised to give better agreement with recent laboratory and
  theoretical work. PAH cross-sections of the emission bands are
  calibrated to match observations in different radiation
  environments.  We have found that in harsh environments, such as in
  starbursts, the integrated cross-section of the C=C bands at 6.2 and
  7.7\,$\mu$m are a factor $\sim 2$ larger, and for the C--H bands at
  11.3 and 12.7\,$\mu$m a factor $\sim 5$ weaker than in the neutral
  regions of the ISM. The cross sections near OB stars are similar to
  the ones derived for starbursts.  

{\bf 3)} With the FORS instrument of the VLT, we have obtained new
  ultra-high signal-to-noise linear and circular spectro-polarimetric
  observations for a selected sample of sight lines.  We have
  performed a detailed study of the instrumental polarisation in an
  attempt to achieve the highest possible accuracy. We show that
  circular polarisation provides a diagnostic on grain shape and
  elongation. However, it is beyond the limit of the observations that
  we have obtained.

  {\bf 4)} The dust model reproduces extinction, linear and circular
  polarisation curves and emission spectra of the diffuse ISM. It is
  set up to keep the number of key parameters to a minimum
  (Table~\ref{para.tab}).  The model accounts for IR observations at
  high galactic latitudes. It can be taken as representative of the
  local dust in the solar neighbourhood.

{\bf 5)} We have applied our dust model on individual sight lines
ranging between $1.2\la A_{\rm V}/{\rm {mag}} \la 2.4$ and $2.9 \la
R_{\rm V} \la 4.6$, towards the early type stars: HD\,37061,
HD\,37903, HD\,93250, and HD\,99872. For these stars we present
polarisation spectra and measure a maximum polarisation of 1 -- 3\,\%.
The IR emission of the star HD\,37903 that is heating the reflection
nebula NGC\,2023 is computed with a radiative transfer program
assuming spherical symmetry.  In the Spitzer/IRS spectrum of the
massive star HD\,99872 we detect an excess emission over its photosphere
that is steeply rising towards longer wavelengths and pointing towards
a cool dust component.

  {\bf 6)} Linear polarisation depends on the type of the spheroid,
  prolate or oblate, its elongation, and the alignment efficiency. We
  have found that the spectral shape of the polarisation is critically
  influenced by the assumed lower and upper radius of dust that is
  aligned. In conclusion polarisation helps to determine the otherwise
  purely constrained upper size limit of the dust particles. The
  observed linear polarisation spectra are better fit by prolate than
  by oblate grains.  For accounting of the polarisation typically only
  silicates, with an elongation of about 2 and radii between 100\,nm
  and 500\,nm need to be aligned.


\begin{acknowledgements} {We are grateful to Endrik Kr\"ugel for
    helpful discussions and thank Gaicomo Mulas for providing their
    PAH cross sections in electronic form. NVV was partly supported by
    the RFBR grant 13-02-00138. This work is based on observations
    collected at the European Southern Observatory, VLT programs
    386.C-0104. This research has made use of the SIMBAD database,
    operated at CDS, Strasbourg, France. This work is based on data
    products of the following observatories: Spitzer Space Telescope,
    which is operated by the Jet Propulsion Laboratory, California
    Institute of Technology under a contract with NASA. Infrared Space
    Observatory funded by ESA and its member states.  Two Micron All
    Sky Survey, which is a joint project of the University of
    Massachusetts and the Infrared Processing and Analysis
    Center/California Institute of Technology, funded by the National
    Aeronautics and Space Administration and the National Science
    Foundation. Wide-field Infrared Survey Explorer, which is a joint
    project of the University of California, Los Angeles, and the Jet
    Propulsion Laboratory/California Institute of Technology, funded
    by the National Aeronautics and Space Administration.  Data
    available at the Space Astronomy Laboratory (SAL), a unit of the
    Astronomy Department at the University of Wisconsin. }
\end{acknowledgements}

\end{document}